\documentclass[aps,twocolumn,prd,showpacs,showkeys,preprintnumbers,superscriptaddress,nobibnotes,floatfix,longbibliography,nofootinbib]{revtex4-1}
\pdfoutput=1
\usepackage[english]{babel}
\usepackage{amsmath,amssymb,amsbsy,amstext,amsthm,booktabs,exscale,relsize,slashed,graphicx,amsfonts,upgreek,xcolor,amsmath}
\usepackage{multirow,dcolumn,bm,enumerate,url,hyperref}
\usepackage[capitalise]{cleveref}
\usepackage{fontawesome}
\usepackage{xspace}
\usepackage[normalem]{ulem}
\definecolor{lightsabergreen}{rgb}{.14,.64,.14}
\definecolor{lightgreen}{rgb}{.14,.44,.14}

\hypersetup{
  colorlinks   = true, 
  urlcolor     = lightsabergreen, 
  linkcolor    = lightsabergreen, 
  citecolor   = lightsabergreen 
}

\newcommand{\gev}{{\rm GeV}}

\newcommand{\diff}{\textrm{d}}
\newcommand{\geom}{\mathrm{geom}}
\newcommand{\RE}{{R_\oplus}}
\newcommand{\Ratm}{R_{\oplus,\rm atm}}
\newcommand{\Nuc}[2]{\ensuremath{^{#2}\mbox{#1}}}

\newcommand{\et}{{\it et al.}}
\newcommand{\ie}{{\it i.e.}}
\newcommand{\uvec}[1]{\hat{#1}}

\newcommand{\githubmaster}{\href{https://github.com/songningqiang/DaMaSCUS-EarthCapture}{
\faGithub}\xspace}

\begin{document}

\preprint{
\begin{minipage}{5cm}
\small
\flushright
UCI-HEP-TR-2022-18
\end{minipage}} 

\count\footins = 1000
\title{\bf Light Dark Matter Accumulating in Planets: Nuclear Scattering}

 \author{Joseph Bramante}
 \thanks{{\scriptsize Email}: \href{mailto:joseph.bramante@queensu.ca}{joseph.bramante@queensu.ca}; {\scriptsize ORCID}: \href{http://orcid.org/0000-0001-8905-1960}{0000-0001-8905-1960}}
 \affiliation{\smaller The Arthur B. McDonald Canadian Astroparticle Physics Research Institute, \protect\\ Department of Physics, Engineering Physics, and Astronomy, \protect\\ Queen's University, Kingston, Ontario, K7L 2S8, Canada}
 \affiliation{\smaller Perimeter Institute for Theoretical Physics, Waterloo, ON N2J 2W9, Canada}

\author{Jason Kumar}
 \thanks{{\scriptsize Email}: \href{mailto:jkumar@hawaii.edu}{jkumar@hawaii.edu}}
\affiliation{\smaller Department of Physics and Astronomy, University of Hawai'i, Honolulu, HI 96822, USA}

 \author{Gopolang Mohlabeng}
  \thanks{{\scriptsize Email}: \href{mailto:gmohlabe@uci.edu}{gmohlabe@uci.edu}}
 \affiliation{\smaller The Arthur B. McDonald Canadian Astroparticle Physics Research Institute, \protect\\ Department of Physics, Engineering Physics, and Astronomy, \protect\\ Queen's University, Kingston, Ontario, K7L 2S8, Canada}
\affiliation{\smaller Perimeter Institute for Theoretical Physics, Waterloo, ON N2J 2W9, Canada}
\affiliation{\smaller Department of Physics and Astronomy, University of California, Irvine, CA 92697, USA}
\affiliation{\smaller Department of Physics, Simon Fraser University, Burnaby, BC, V5A 1S6, Canada}

\author{Nirmal Raj}
\thanks{{\scriptsize Email}: \href{mailto:nraj@iisc.ac.in}{nraj@iisc.ac.in}}
\affiliation{\smaller TRIUMF, 4004 Wesbrook Mall, Vancouver, BC V6T 2A3, Canada}
\affiliation{\smaller Centre for High Energy Physics, Indian Institute of Science, C. V. Raman Avenue, Bengaluru 560012, India}

\author{Ningqiang Song}
\thanks{{\scriptsize Email}: \href{mailto:songnq@itp.ac.cn}{songnq@itp.ac.cn}}
\affiliation{\smaller Department of Mathematical Sciences,\protect \\ University of Liverpool,  Liverpool, L69 7ZL, United Kingdom}
\affiliation{\smaller The Arthur B. McDonald Canadian Astroparticle Physics Research Institute, \protect\\ Department of Physics, Engineering Physics, and Astronomy, \protect\\ Queen's University, Kingston, Ontario, K7L 2S8, Canada}
\affiliation{\smaller Institute of Theoretical Physics, Chinese Academy of Sciences, Beijing, 100190, China}

\begin{abstract}

We present, for the first time, a complete treatment of strongly-interacting dark matter capture in planets, taking Earth as an example. We focus on light dark matter and the heating of Earth by dark matter annihilation, addressing a number of crucial dynamical processes which have been overlooked, such as the ``ping-pong effect" during dark matter capture.
We perform full Monte-Carlo simulations and obtain improved bounds on strongly-interacting dark matter from Earth heating and direct detection experiments for both spin-independent and spin-dependent interactions, while also allowing for the interacting species to make up a sub-component of the cosmological dark matter. \githubmaster

\end{abstract}
\maketitle

\section{Introduction \label{sec:intro}}

Celestial bodies are well motivated laboratories to search for dark matter (DM) due to their ability to capture DM particles and enhance the flux of their annihilation products.
The standard framework for studying DM capture and evaporation in a variety of astrophysical objects was largely established  by Refs.~\cite{Gould:1987ju,Gould:1987ir,Gould:1987ww, Zentner:2009is,Kong:2014mia, Alhazmi:2016qcs, Feng:2016ijc, Garani:2017jcj, Feng:2015hja,Baum:2016oow, Catena:2016kro, Adler:2008ky, Iocco:2008xb, Guver:2012ba,Baryakhtar:2017dbj, Raj:2017wrv, Joglekar:2019vzy, Acevedo:2019agu, Joglekar:2020liw, Leane:2020wob, Dasgupta:2020dik, Garani:2020wge, Leane:2021tjj, Leane:2021ihh, Lopes:2021jcy, Garani:2021feo, Bell:2021fye, Banks:2021sba, Bramante:2021dyx, Coffey:2022eav,Maity:2021fxw,Gould:1989tu,Gould:1991va}, 
in part because the annihilation of DM in these systems can yield signals of weakly coupled particles which can be observed by large volume experiments on Earth and various other space-based detectors.

The focus of this work is on revisiting DM capture and evaporation in the Earth in the optically thick regime, where dark matter is expected to be slowed down markedly and thermalize in the overburden before reaching deep underground direct detection experiments~\cite{Xia:2021vbz,Xia:2022tid}, yielding open parameter space at large couplings, even for relatively small masses. On the other hand, novel detection techniques have been developed in the last few years which are sensitive to low energy deposition down to meV~\cite{Hochberg:2015pha,Hochberg:2015fth,Hochberg:2021pkt,Hochberg:2021ymx,Hochberg:2019cyy,Chiles:2021gxk,Schutz:2016tid,knapen2017light,Caputo:2019cyg,Griffin:2018bjn,Knapen:2017ekk,Cox:2019cod,Sanchez-Martinez:2019bac,Hochberg:2017wce,Geilhufe:2018gry,geilhufe2020dirac,Coskuner:2019odd}, paving the way for the detection of DM thermalized in the Earth crust~\cite{Leane:2022hkk,Das:2022srn}.
These motivate a more detailed study of dark matter accumulated in the Earth.


In this paper, we perform the first proper analysis of dark matter capture in the Earth by using a Monte Carlo (MC) simulation that fully encapsulates the dynamics of light dark matter capture, in which the multiple scattering effect is accurately accounted for. We find that a proper dynamical treatment greatly alters the capture rate of light dark matter, relative to previous studies' simplified assumptions.  To show the salient feature of the analysis, we derive Earth heating constraints on DM models that would cause excess heating of the Earth through DM annihilation to visible matter~\cite{Mack:2007xj, Bramante:2019fhi}. In doing so, we combine capture, annihilation and evaporation processes, with 
important effects arising from multiple scattering implemented throughout.

We also present the first comprehensive analysis of current direct detection constraints on spin-dependent (SD) dark matter interactions. We substantially improve the previous conservative constraints where only DM particles unscattered before reaching the detector were considered~\cite{Hooper:2018bfw}, by including the important effects of SD form factors, the angular dependence of DM particle trajectories in the overburden, and the velocity distribution of DM. Confronting these newly derived limits, we find Earth heating excludes a wide range of new parameter space in the strongly interacting regime that is not excluded by existing direct detection experiments.

The remainder of this paper is structured as follows. In Section \ref{sec:captureMC} we compute the capture rate of dark matter from MC simulations. Then in Section \ref{sec:DM_evapmain} we describe the distribution of dark matter number density throughout the Earth and evaluate the evaporation rate. Section \ref{sec:Earthheting} is devoted to the Earth heating from dark matter annihilation. Finally, in Section \ref{sec:results} we conclude by showing the cross section limits from Earth heating and direct detection experiments.  Technical details are elaborated on in the Appendices: In Appendix~\ref{app:MC} we provide more details on the MC. Comparison of the MC results against the single-scatter capture formalism is presented in Appendix~\ref{app:analytical}, with comments on the effects of the solar gravitational potential made in Appendix~\ref{app:solargravity}. Comparision with the multi-scatter formalism is also given in Appendix~\ref{app:multianalytical}. The resonant capture behavior found by MC is explained in Appendix~\ref{app:resonantcapture}. Complete capture results from MC are exhibited in Appendix~\ref{app:variousinteractions}. Details of the dark matter evaporation can be found in Appendix~\ref{app:DM_evap} and the derivation of existing constraints from direct detection experiments are provided in Appendix~\ref{app:Directdetection}.

\section{Dark Matter Capture} 
\label{sec:captureMC}

A DM particle in the halo may scatter with the nuclei in the Earth, losing part of its kinetic energy. If the particle falls below the Earth's escape velocity $v_{\rm esc}=11.2$~km/s, it will be gravitationally captured. In case DM is much heavier than Earth nuclei, then a scatter results in a loss of energy, but a small change in direction.
In the limit of many scatters, DM is essentially guaranteed to be captured, leading to the geometric capture rate $C_\oplus^{\rm geom}$ at strong coupling, where all DM particles that bombard the Earth are captured.  But if DM is much lighter than the nuclei it scatters against, then the DM particle's direction is expected to be essentially randomized after every 
scatter, much like a ping-pong ball scattering off a bowling ball.  In that case, even if dark matter couples arbitrarily strongly to nuclei, it need not scatter more than once, as it may be reflected away at the first few scatters.  This would lead to a sharp suppression in the capture rate. 

To study the multiple scattering effects in dark matter capture, we use the \texttt{DaMaSCUS-EarthCapture} code, developed from the \texttt{DaMaSCUS} code~\cite{Emken:2017qmp,emken2017damascus}.  We improve the code in various ways: 1) We add the crust layer and the atmosphere of the Earth and their chemical compositions, allowing dark matter to scatter and stop there.  2) We consider the acceleration of halo DM particle  by the Sun's and Earth's gravitational potential. The halo DM velocity $u_\chi$ is drawn from the Maxwellian distribution translated to the Earth frame. Upon arriving at the Earth, DM is assigned a velocity $w=(u_\chi^2+v_s^2+v_{\rm esc}^2)^{1/2}$, where $v_s=42.2$~km/s is the the escape velocity from the Sun at 1 AU . 3) If DM leaves the Earth with a velocity $v \leq v_{\rm esc}$, it reenters the Earth at the exit point with the opposite velocity, as these dark matter particles will follow an elliptical path and reenter the Earth at some point due to gravity. We neglect the thermal motion of Earth nuclei, and simulate the trajectories of DM passing through the Earth, either streaming freely or scattering with the nuclei. This is justified, as the thermal velocity of Earth nuclei is much smaller than the velocity of DM in the halo. The probability of DM scattering after traveling freely over a length $L$ is $P=1-\exp(-\int_0^L dx/\lambda(x))$, with $\lambda$ the mean free path of DM in Earth. The scattering angle is randomized between 0 and $\pi$ in the center-of-mass frame for velocity and momentum transfer independent scattering, which is then translated to the Earth frame.
A DM particle is considered lost when it leaves the Earth with a velocity $v>v_{\rm esc}$, and captured when it reaches a velocity $v<v_{\rm esc}$ inside the Earth. We then determine the capture fraction, $f_C$, which is defined as the fraction of dark matter particles reaching the surface of the Earth and subsequently captured. The MC capture rates are computed with $C_\oplus=f_CC_\oplus^{\rm geom}$. Additional details of the implementation of the MC are presented in Appendix~\ref{app:MC}.

\begin{figure}
    \centering
    \includegraphics[width=\columnwidth]{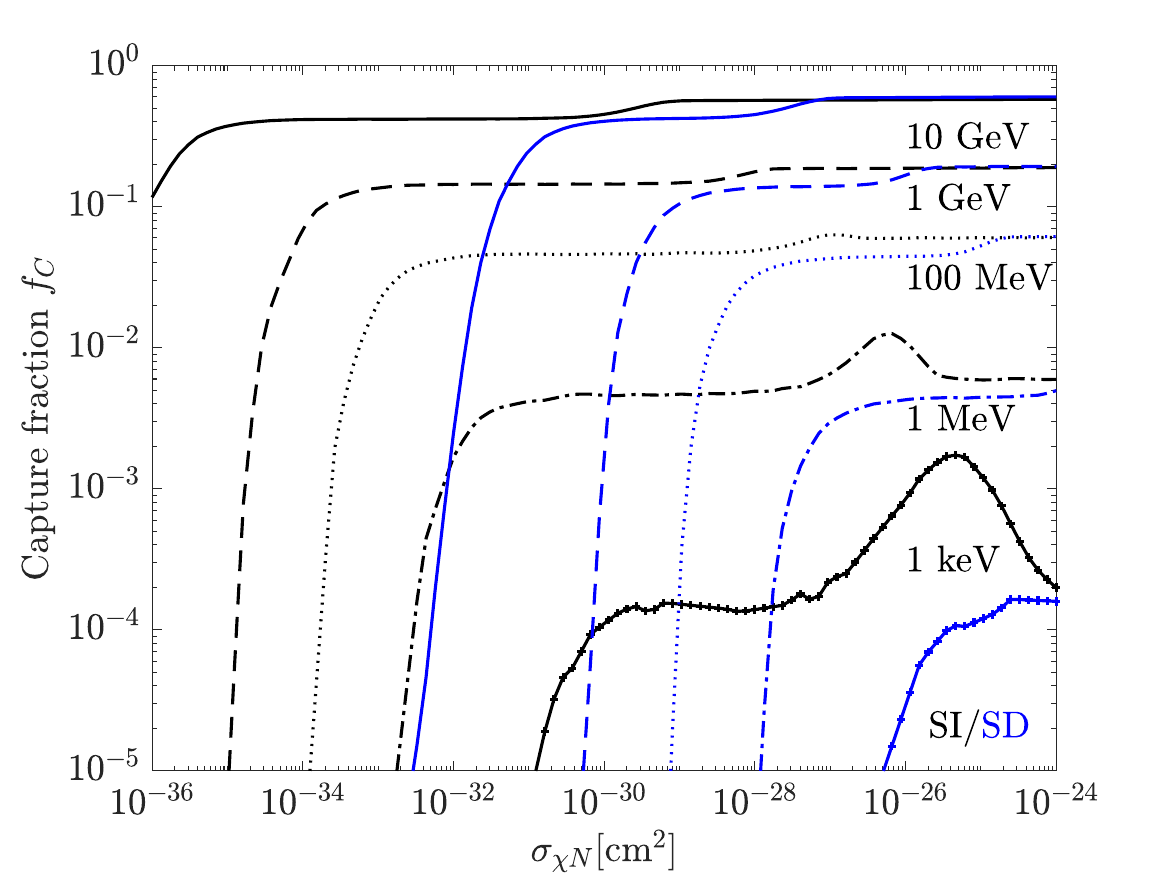}    
    \caption{Dark matter capture in the Earth  from Monte Carlo simulation using \texttt{DaMaSCUS-EarthCapture}. The black and blue lines depict the fraction of dark matter particles that are captured among those that impinge on the Earth as a function of spin-independent (SI) and spin-dependent (SD) ($a_p=a_n=1$) dark matter nucleon scattering cross sections respectively, for the various dark matter masses indicated by different line styles.}
    \label{fig:capMC}
\end{figure}

We consider two types of interactions, spin-independent (SI) and spin-dependent (SD) nuclear scattering 
(both taken to be velocity-independent). 
In the former case, the nuclear scattering cross section at zero momentum transfer is given by
\begin{equation}
    \sigma_{j,0}^{\rm SI}=\left(\dfrac{\mu_{A_j}}{\mu_N}\right)^2 A_j^2\sigma^{\rm SI}_{\chi N}\,,
    \label{eq:xsnucleusSI}
\end{equation}
where $A_j$ is the mass number of a Standard Model (SM) nucleus, $\mu_{A_j}$ ($\mu_N$) is the reduced mass between DM and the nucleus (nucleon), and $\sigma^{\rm SI}_{\chi N}$ is the DM-nucleon SI scattering cross section. For the latter, we have instead
\begin{equation}
    \sigma_{j,0}^{\rm SD}=\left(\dfrac{\mu_{A_j}}{\mu_N}\right)^2 S_{J_j}\left(a_p\langle S_p\rangle+a_n\langle S_n\rangle \right)^2\sigma^{\rm SD}_{\chi N}\,,
    \label{eq:xsnucleusSD}
\end{equation}
where $S_J=4(J+1)/(3J)$ and $J$ is the total nuclear spin, $\sigma^{\rm SD}_{\chi N}$ is the DM-nucleon SD scattering cross section, and $\langle S_p\rangle$ and $\langle S_n\rangle$ represent the average spin of protons and neutrons in a nucleus, respectively. We also explore three scenarios, isospin-independent scattering $a_p=a_n=1$, proton-only scattering $a_p=1$, $a_n=0$ and neutron-only scattering $a_p=0$, $a_n=1$. We will set the momentum-dependent form factor to unity unless explicitly stated, the inclusion of which turns out not to change the results significantly.

We show the capture fraction of DM in Figure~\ref{fig:capMC}. For relatively small cross sections $\sigma_{\chi N}^{\rm SI}\lesssim 10^{-33}$~cm$^2$, the capture fraction becomes suppressed when the DM mass is much smaller than the target nuclei $m_\chi\ll m_A$, as the energy transfer $\propto q^2/m_A$ in a typical scatter becomes small 
(where $q\propto m_\chi$ is the momentum transfer), implying that more scatters are necessary before a DM particle can be 
captured. In most Earth layers, O or N is the dominant scattering target and is relatively light and abundant, except in the core, where Fe has the largest mass fraction. This suppression tends to get milder as the cross section increases, since DM may scatter more often as it crosses the Earth. 
At $\sigma_{\chi N}^{\rm SI}\gtrsim10^{-30}$~cm$^2$, the capture saturates down to keV DM mass, \ie~increasing the cross section further contributes little to the capture fraction.
However, note that for $m_\chi \ll m_A$, the $f_C$ saturates to a value which is $\ll 1$.  
A numerical fit suggests that the saturation capture fraction $f_C\propto \sqrt{m_\chi}$, which is consistent with the fraction of reflected dark matter in analytical treatment. We also notice that the capture fraction of light dark matter
($m_\chi\ll m_A$) is peaked near $\sigma_{\chi N}^{\rm SI} \gtrsim 10^{-26}$~cm$^2$.  
This is due to the summation of possible dark matter trajectories as discussed in Appendix~\ref{app:resonantcapture}. On the other hand, the capture fraction is close to 100\% for $m_\chi \gtrsim 10~\gev$ when $\sigma_{\chi N}^{\rm SI} \gtrsim 10^{-35}$~cm$^2$. 
For large scattering cross sections, dark matter capture is dominated by scattering with N in the atmosphere.  For smaller 
cross sections, capture is instead dominated by scattering with O, Si, or Fe in the crust and mantle.  For even smaller cross sections, 
capture is dominated by scattering with Fe in the core.
Trends similar to those discussed above for SI interactions also apply to SD interactions, with DM scattering with nuclei with non-zero proton or neutron spins, but for much higher dark matter-nucleon scattering cross sections. Full results of dark matter capture with different types of interactions are presented in Appendix~\ref{app:variousinteractions}.

For comparison, we also show the analytical treatment of dark matter capture due to single scatter in Appendix~\ref{app:analytical} and multiple scatters in Appendix~\ref{app:multianalytical}. We find that MC typically provides more exact descriptions of the capture rate, as the full kinematical properties of the capture processes are addressed. We also stress that the acceleration of dark matter due to the solar gravitational potential is usually neglected. Although this does not change the MC capture rates, it may significantly affect the results considering only single scatter. This effect is discussed in Appendix~\ref{app:solargravity}.

\section{Dark Matter Evaporation \label{sec:DM_evapmain}}

 In the optically thick regime where the cross section is large (for SI (SD) this corresponds to $\sigma_{\chi N}^{\rm SI}\gtrsim 10^{-36}$~cm$^2$ ($\sigma_{\chi N}^{\rm SD}\gtrsim 10^{-32}$~cm$^2$)) where the Knudsen number $K\lesssim 1$, dark matter particles may thermalize with their local ambient environment due to frequent scattering momentum exchange. 
In this case, we may take the DM to be in local thermal equilibrium (LTE); essentially, we may assume that, in every 
small volume element, DM is an ideal gas in thermal equilibrium with SM matter at temperature $T(r)$ at radius $r$, and in 
diffusive equilibrium with DM in the surrounding volume. For relatively light DM ($m_\chi\lesssim 0.1$~GeV), captured DM dwells towards the surface of the Earth due to the temperature gradient in the upper mantle and the crust. Heavy DM particles tend to sink down.

 Due to the thermal motion of Earth nuclei, DM may scatter with a nucleus and acquires a high enough velocity to escape from the Earth.
The corresponding evaporation rate is described as~\cite{Garani:2017jcj}
\begin{equation}
\begin{split}
    E_\oplus &= \sum\limits_j\int_0^{\Ratm} 4\pi r^2 n_\chi(r)s(r)dr\times\\
    &\int_0^{v_e(r)}4\pi u_\chi^2f_\oplus du_\chi\int_{v_e(r)}^\infty R_j^+(u_\chi\rightarrow v)dv\,,
\end{split}    
    \label{eq:evaprates}
\end{equation}
where $f_\oplus$ is the thermal distribution of captured DM, $n_\chi(r)$ is the LTE density profile, $v_e(r)$ is the Earth escape velocity at radius $r$  and $R_j^+$ describes the scattering effects. The integral is carried out through the Earth including the atmosphere and $\Ratm=6471$~km.  Note that in the optically thick regime, DM particles with a velocity above the escape velocity may not actually evaporate, as they may scatter with the Earth matter several times before making their way out. This effect is encapsulated in the $s(r)$ factor. Since the total evaporation rate depends on the total number of captured DM $N_C=\int 4\pi n_\chi(r')r'^2dr'$ which is yet to be solved for, we define the evaporation rate per particle $\title{E_\oplus}\equiv E_\oplus/N_C$, which now depends only on the DM mass and scattering cross section. We find that DM with mass $m_\chi\gtrsim 10$~GeV can hardly escape from the Earth regardless of the cross section, while evaporation is significant for $m_\chi\ll m_A$. More details on dark matter evaporation are found in Appendix~\ref{app:DM_evap}.

\section{Earth Heating}
\label{sec:Earthheting}

The DM depletion rate due to annihilation, normalized by the total number of DM particles, is given by
\begin{equation}
    A_\oplus=\dfrac{\langle \sigma v\rangle_{\chi\chi}}{N_C^2}\int_0^{\Ratm} n_\chi^2 4\pi r^2dr\,,
    \label{eq:normannihilation}
\end{equation}
where we assume the thermal averaged  s-wave annihilation cross section $\langle \sigma v\rangle_{\chi\chi}\simeq 3\times 10^{-26}$~cm$^3$/s. The rate of change of the number of DM particles captured within the Earth is then expressed as 
\begin{equation}
\frac{dN_C}{dt} = C_\oplus - \left(\frac{E_\oplus}{N_C} \right) N_C  - 
A_\oplus N_C^2\,,
\label{eq:NCequation}
\end{equation}
assuming the DM particle is its own anti-particle.  In the equilibrium limit, we find $dN_C / dt =0$.  But more generally, we  solve Eq.~\eqref{eq:NCequation} for $N_C (t)$, and evaluate the total annihilation rate, 
\begin{equation}
    \Gamma_\oplus (t)=(1/2) A_\oplus N_C(t)^2\,,
    \label{eq:totalannrate}
\end{equation}
at present times.

\begin{figure*}
    \centering
    \includegraphics[width=0.48\textwidth]{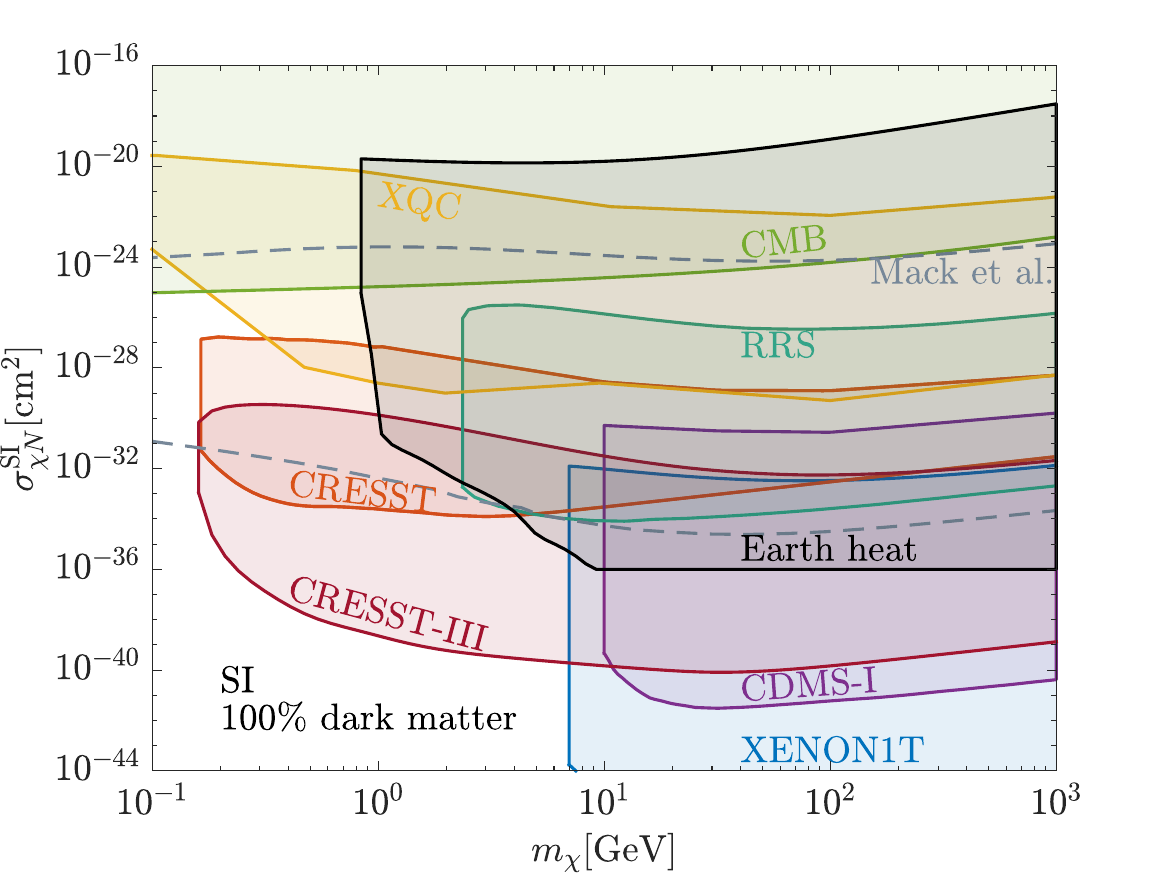}
    \includegraphics[width=0.48\textwidth]{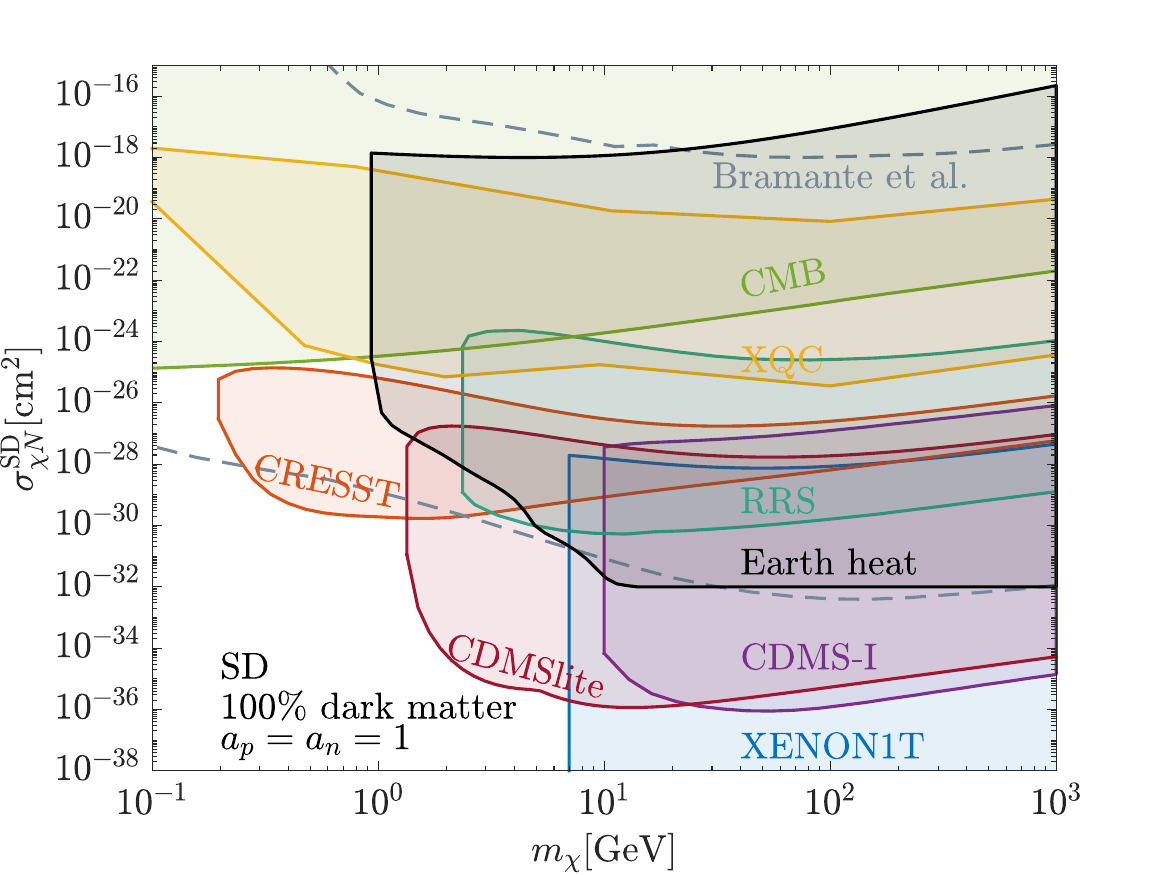}
    \includegraphics[width=0.48\textwidth]{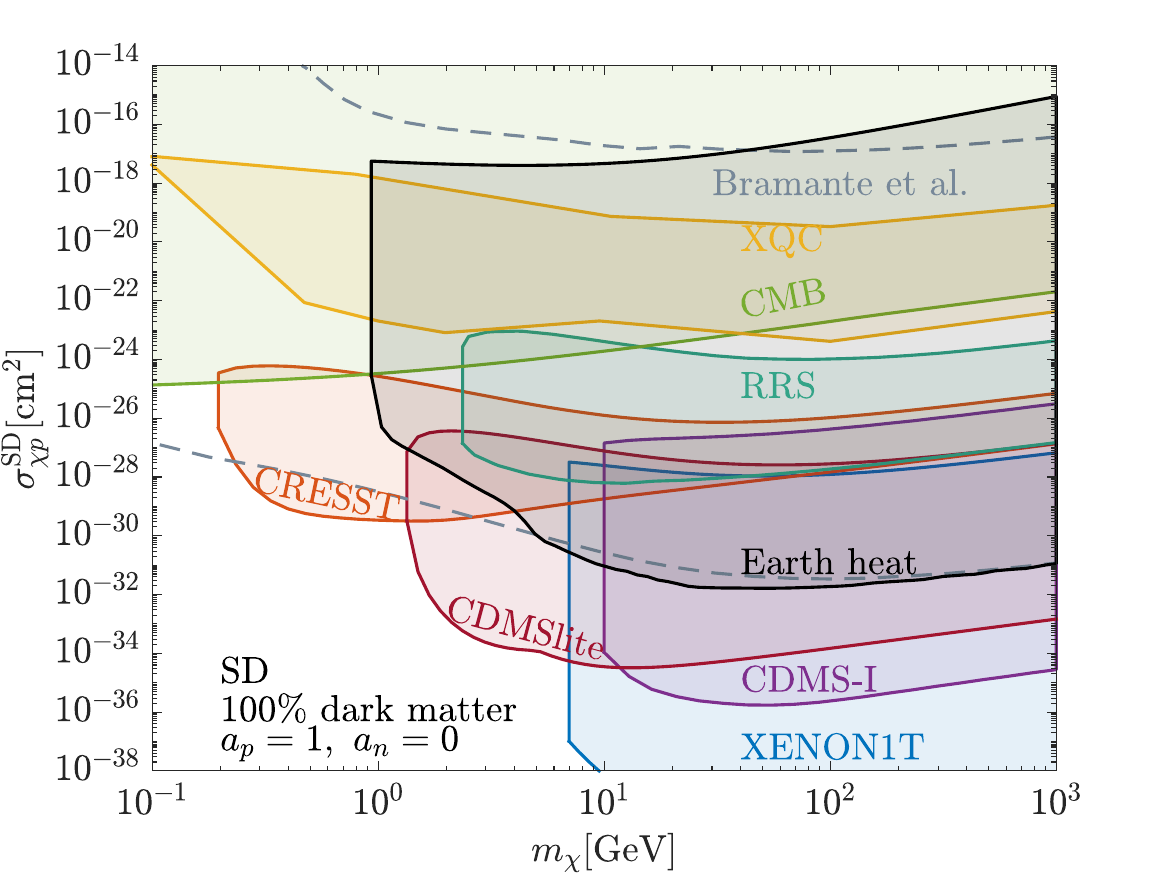}
    \includegraphics[width=0.48\textwidth]{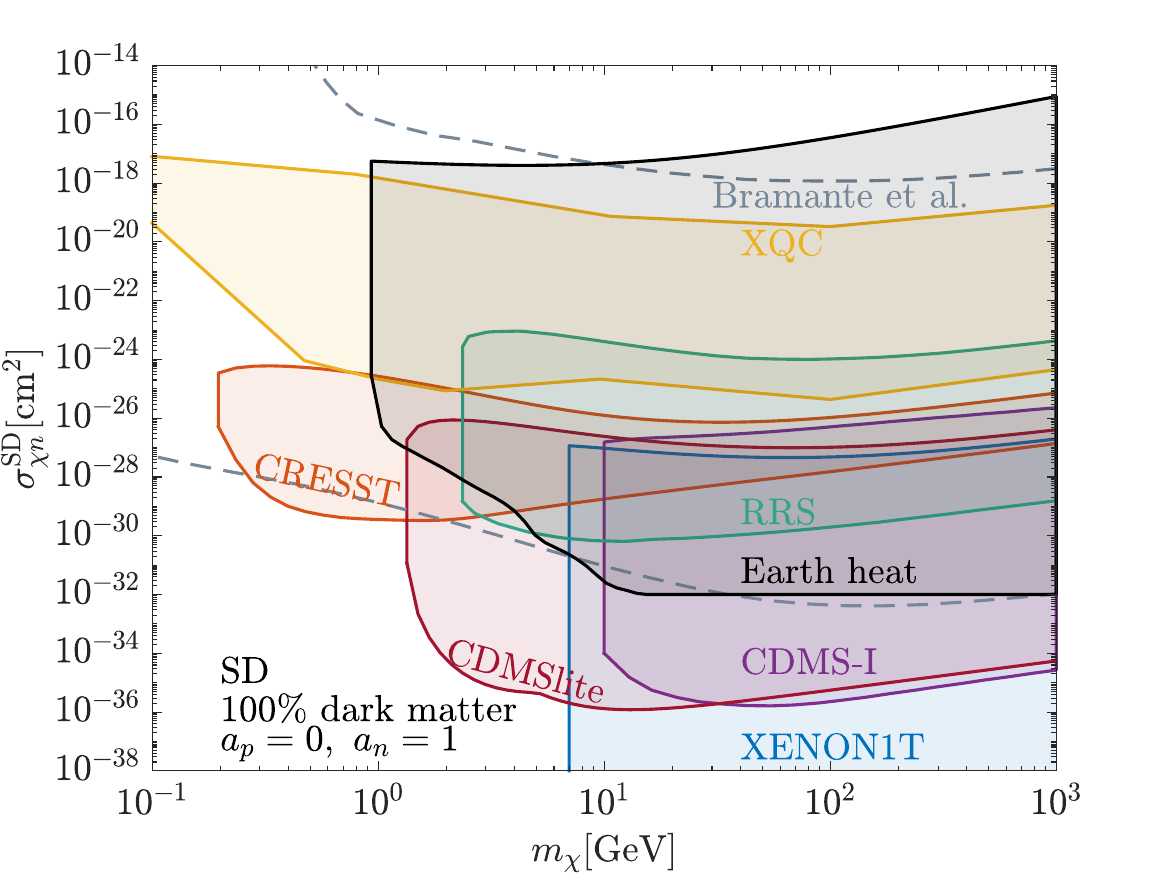}       
    \caption{Earth heating limit from dark matter annihilation for spin-independent (SI) and spin-dependent (SD) nuclear scattering interactions. Also shown are results from  CMB~\cite{Dvorkin:2013cea,Gluscevic:2017ywp}, XQC~\cite{Erickcek:2007jv}, RRS~\cite{rich1987search}, CRESST 2017 surface run~\cite{CRESST:2017ues}, CDMS-I~\cite{CDMS:2002moo}, CDMSlite~\cite{SuperCDMS:2017nns}, CRESST-III~\cite{CRESST:2019jnq} and XENON1T~\cite{XENON:2018voc}. See text for details. $\chi$ is assumed to constitute 100\% of the dark matter. {\it Upper Left}: SI interaction. The region enclosed by the dashed gray lines is the Earth heating limit from Mack {\it et. al.}~\cite{Mack:2007xj}. {\it Upper Right:} SD interaction with isospin-independent nuclear response $a_p=a_n=1$. Dashed gray lines display the Earth heat limit from Bramante {\it et.~al.}~\cite{Bramante:2019fhi}. {\it Lower Left:} SD interaction with proton only scattering. {\it Lower Right:} SD interaction with neutron only scattering.}
    \label{fig:Earthheatlimit}
\end{figure*}

Bounds on Earth heating by DM are obtained by requiring that DM annihilating inside the Earth cannot contribute more thermal energy than that observed flowing out of the Earth's surface. 
It is well established that the heat presently flowing from the Earth is less than $\sim$44~TW \cite{williams1974, lister1990, JAUPART2015223, Davies1980, Davies1980_2, sclater1980, pollack1993, davies2010}, where a substantial portion of this heat flow is attributable to the decay of potassium and uranium \cite{arevalo2009}. Complementary to the Earth heat measurement, recent observation of geoneutrinos at Borexino shows the Earth’s radiogenic heat is $38.2_{-12.7}^{+13.6}$~TW~\cite{Borexino:2015ucj,Kumaran:2021lvv}, while the neutrino measurement at KamLAND indicates a lower radiogenic heat of around 14.6~TW~\cite{abe2022abundances}. We conservatively require the heat flow from DM annihilation to be no more than 44~TW, but note that a better determination of the radiogenic heat may substantially improve this limit. Since the temperature profile of the Earth should be mildly different at early geological times, we evaluate the number of DM trapped in the Earth by solving Eq.~\eqref{eq:NCequation} after $t=10^9$ years. This is conservative in the sense that additional heating of the Earth over geological timescales should only increase the predicted present-day heat flowing from the Earth's surface. The total annihilation rate is obtained using Eq.~\eqref{eq:totalannrate}, which, multiplied by $2m_\chi$, yields the heat flow powered by DM annihilation. We always assume DM annihilates to visible final states, and the annihilation deposits 100\% of its mass-energy in the form of heat.

To determine the maximum cross-section for which the Earth heating bound is applicable it is necessary to consider the maximum cross-section for which DM annihilation occurs predominantly \emph{below} the surface of the Earth. For a large enough DM-nucleon scattering cross-section, DM can drift slowly enough through the Earth's atmosphere that it annihilates predominantly within the atmosphere before reaching the Earth's surface.
 We obtain a conservative upper limit on the Earth heating bound cross-section by requiring that no more than 10\% of the DM captured annihilates on its way to the Earth's surface $\Gamma_{\oplus,\rm atm} < C_\oplus /10$. To determine this condition, we sum over annihilation in atmospheric shells of one kilometer thickness,
\begin{equation}
  \Gamma_{\oplus,\rm atm}  = \sum_{r_i=R_\oplus}^{R_{\oplus,\rm atm}} \frac{\langle \sigma v\rangle_{\chi\chi} \Big(C_\oplus t_{\rm drift}(r_i,r_{i+1}) \Big)^2}{V_{\rm shell}},
\end{equation}
where the volume of each shell is simply $V_{\rm shell} = \frac{4 \pi}{3} (r_{i+1}^3-r_i^3)$. The time for the DM to drift within a shell is~\cite{Acevedo:2020gro}
\begin{equation}
   t_{\rm drift}(r_i,r_{i+1}) = 
\sum_j \frac{\sigma_j}{G m_\chi}   \int_{r_i}^{r_{i+1}} \frac{n_j \sqrt{3 m_{A_j} T}}{M(r)/r^2}  \diff r  \,,
\label{eq:tdrift}
\end{equation}
where $\sigma_j$ is the dark matter nuclear scattering cross section which we approximate by $\sigma_{j,0}$ in Eq.~\eqref{eq:xsnucleusSI} or Eq.~\eqref{eq:xsnucleusSD} as appropriate, and we sum over all nuclear targets in the Earth atmosphere.

\section{Results and Discussions}
\label{sec:results}

\begin{figure*}
    \centering
    \includegraphics[width=0.48\textwidth]{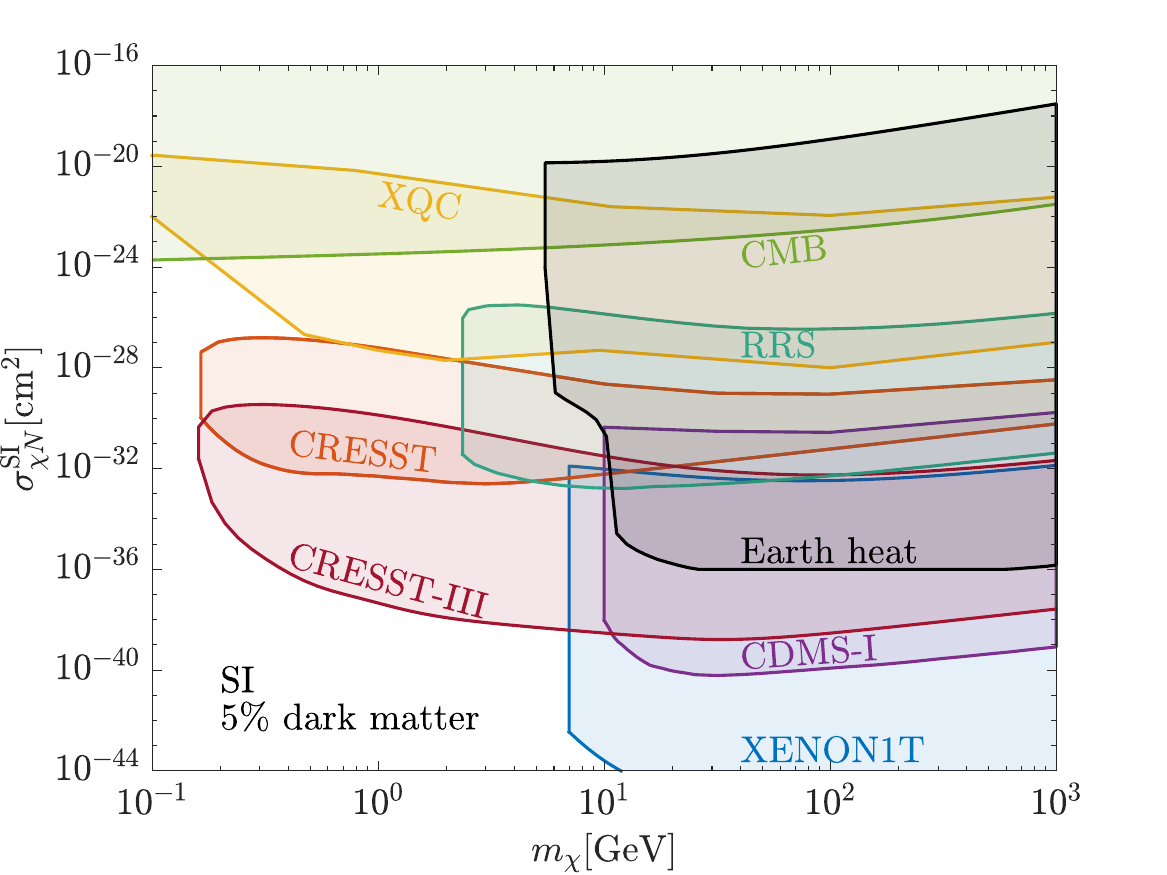}
    \includegraphics[width=0.48\textwidth]{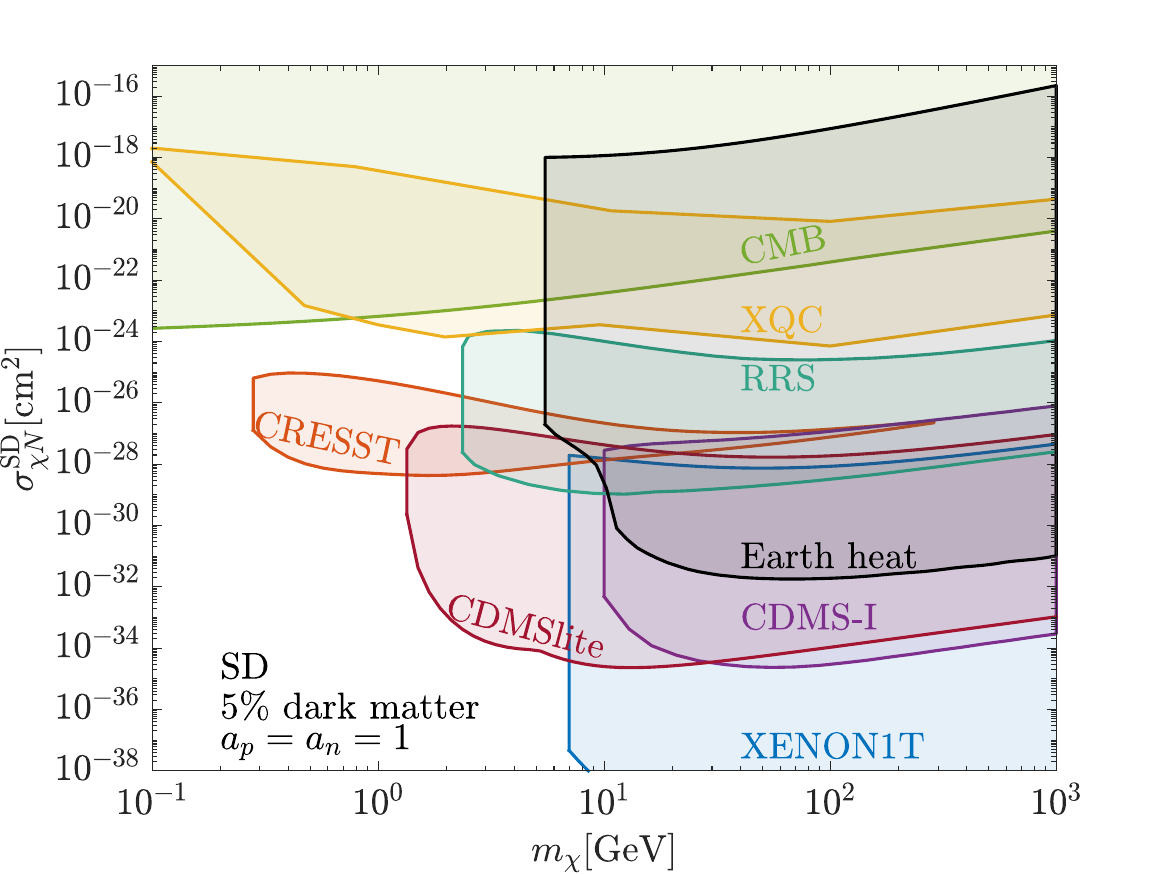}
    \includegraphics[width=0.48\textwidth]{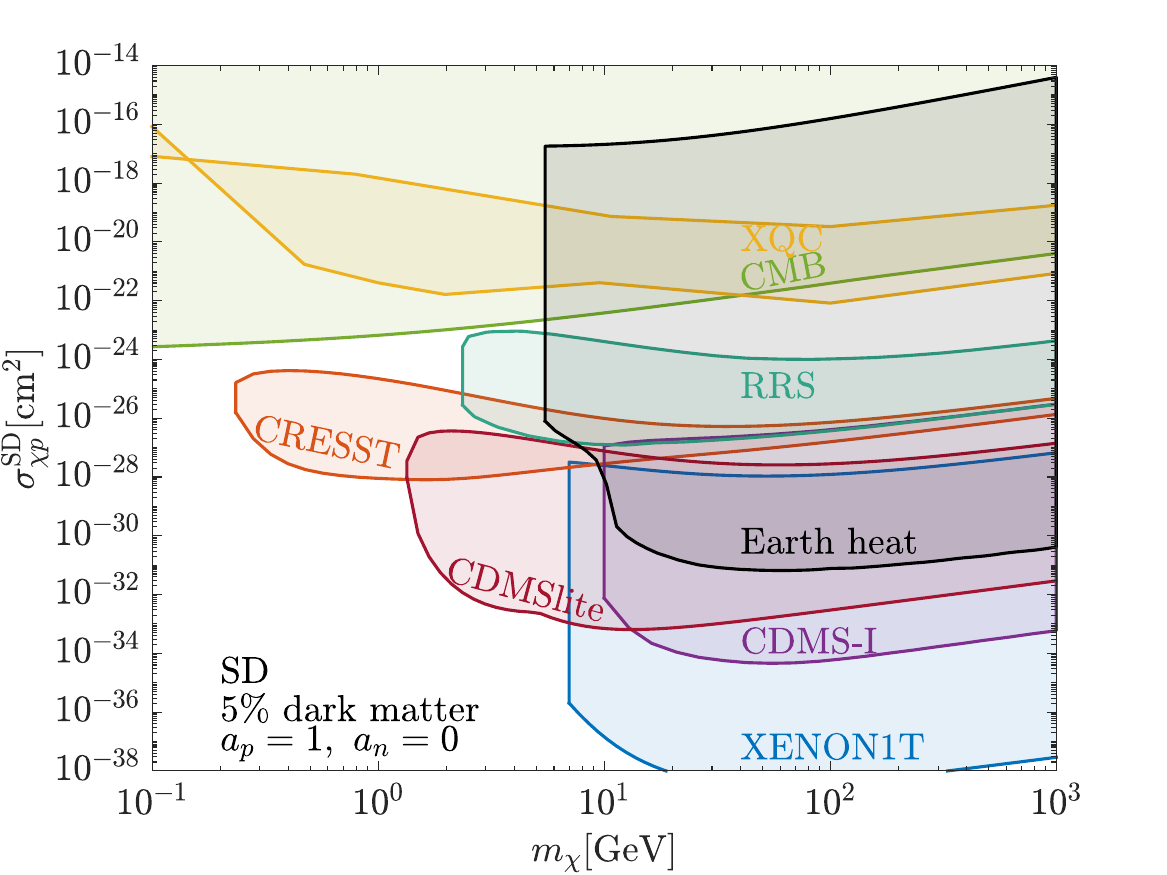} 
    \includegraphics[width=0.48\textwidth]{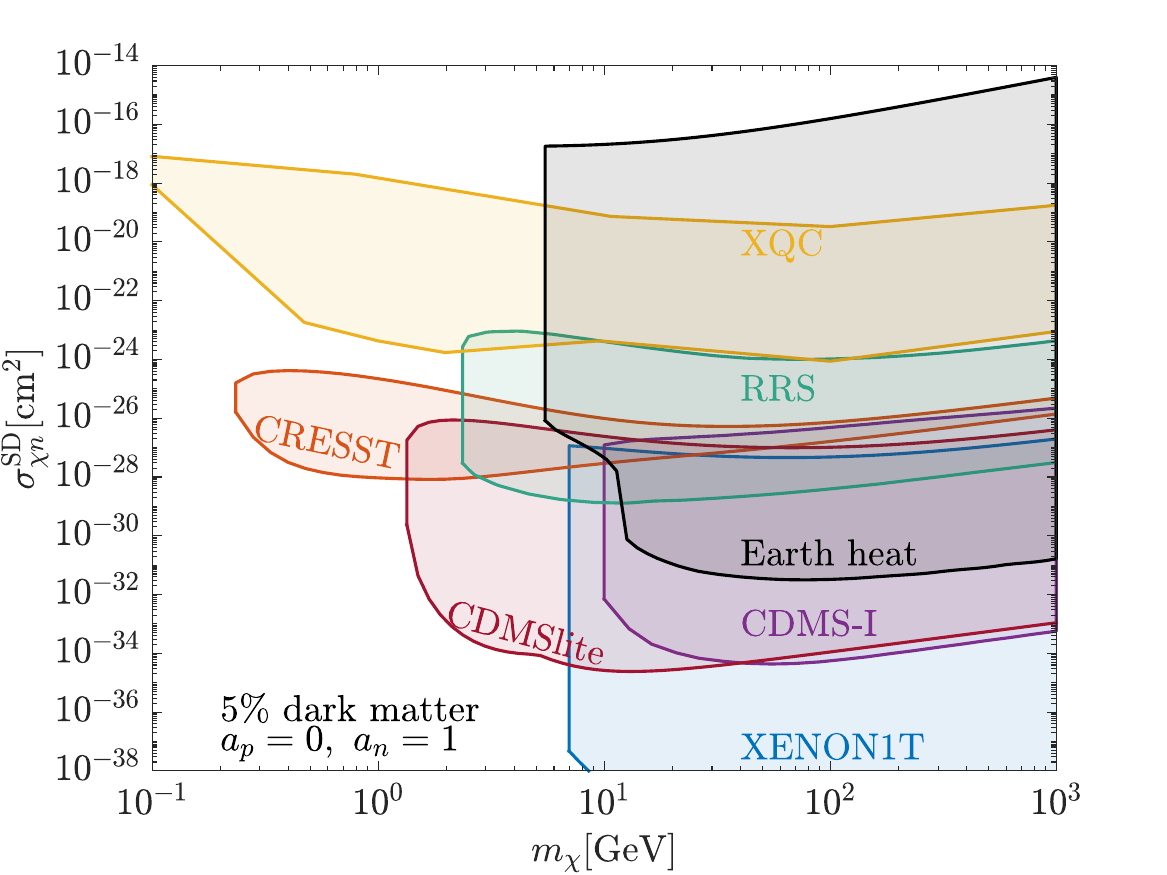}     
    \caption{Earth heating limit from dark matter annihilation for spin-independent (SI) and spin-dependent (SD) nuclear scattering interactions assuming  $\chi$ constitutes 5\% of the cosmological dark matter. {\it Upper Left}: SI interaction.  {\it Upper Right:} SD interaction with isospin-independent nuclear response $a_p=a_n=1$.  {\it Lower Left:} SD interaction with proton only scattering. {\it Lower Right:} SD interaction with neutron only scattering.}
    \label{fig:Earthheatlimit_allSD}
\end{figure*}

We show the Earth heating constraints in Figure~\ref{fig:Earthheatlimit} along with constraints from existing experiments that are mostly newly derived in this work (see also~\cite{Emken:2019tni,Hooper:2018bfw} for relevant discussions). We also allow $\chi$ to be a subdominant component of the cosmological DM and show the corresponding constraints in Figure~\ref{fig:Earthheatlimit_allSD}. The Earth heating limit in this case is obtained by setting the capture rate $C_\oplus=0.05f_CC_\oplus^{\rm geom}$. Assuming thermal freeze-out, we also adopt a correspondingly larger cross section $\langle \sigma v\rangle_{\chi\chi}\simeq 6\times 10^{-25}$~cm$^3$/s. The number of DM is again obtained by solving  Eq.~\eqref{eq:NCequation} with the new capture rate and the total annihilation rate is computed using Eq.~\eqref{eq:totalannrate}. The upper boundary of the exclusion region is derived using the conditions  $\Gamma_{\oplus,\rm atm} < C_\oplus /10$ with the reduced capture rate and enhanced annihilation cross section. The derivation of the constraints from direct detection experiments is presented in Appendix~\ref{app:Directdetection}. 

For SI interaction, the Earth heating places constraints on the DM scattering cross section for $m_\chi>0.84$~GeV (6.2~GeV) assuming 100\% (5\%) cosmological DM in the parameter space where our analysis is valid, \ie~we constrain the cross section $\sigma_{\chi N}^{\rm SI}>10^{-36}$~cm$^2$, where local thermal equilibrium is justified, and $\sigma_{\chi N}^{\rm SI}\lesssim 10^{-20}$~cm$^2$, where DM dominantly drifts down below the surface of the Earth and annihilates there. This effectively cuts the exclusion region of Mack \et~\cite{Mack:2007xj} at lower masses and extends the upper and lower cross section reach. For SD interactions of all types, the lower mass limits shift to 0.93~GeV ($\sim 5.5$~GeV) assuming 100\% (5\%) DM. The excluded cross section is above $10^{-32}$~cm$^2$ to satisfy the local thermal equilibrium distribution, and below about $10^{-17}$~cm$^2$ for DM to drift down. The Earth heating limits exclude a wide range of parameter space that is not covered by direct detection experiments. For SD scattering Earth heating closes the gap between CMB and CRESST in the 0.9~GeV and 2.4~GeV mass window, and the gap between CMB, XQC and RRS above about 10~GeV, particularly for proton-only interaction. Earth heating also excludes the SD neutron-only scattering cross section above the XQC exclusion region. Even more parameter space is precluded if $\chi$ makes up a fraction of the cosmological DM.

We have focused on the accumulation of DM in the Earth for the case in which DM 
can annihilate to visible matter, leading to anomalous heating of the Earth that may be bounded by data. In addition, DM annihilation in the Earth can also produce a flux of neutrinos that may 
be observed at neutrino detectors.  Thus our analysis framework is equally useful for constraining this neutrino flux, which would be an interesting topic of future work. Our work can also be employed to the accumulation of DM particles in the Earth crust, which may facilitate the direct detection of DM in low-threshold detectors. The analysis can also be easily generalized to other astrophysical bodies, including the Sun.

Apart from the velocity-independent SI and SD interactions which we have been considered, dark matter may also scatter 
with SM nucleons through interactions which depend more generally on  momentum transfer, velocity or 
spin, as well as the combination of these~\cite{Fitzpatrick:2012ix,Anand:2013yka,Brod:2017bsw}. Such scenarios were investigated in various direct detection experiments~\cite{CRESST:2018vwt,Kang:2019dbr,DEAP:2020iwi}, where the experimental limits on the scattering cross section were found to be modified as a result of these different interactions~\cite{DEAP:2020iwi}. The effects of momentum and velocity-dependent interactions on dark matter accumulation can be assessed qualitatively. For example, if the cross section scales as a positive power of the relative velocity, 
the evaporation process is more suppressed relative to the capture process.
If the cross-section peaks at low momentum, then low-mass dark matter is more likely to be captured, 
as the small reduced mass 
implies that a larger fraction of final state phase space involves small momentum transfer.
A more sophisticated exploration of these scenarios is left for future work.

\section*{Acknowledgements}

 We thank Rebecca Leane, Juri Smirnov and M.C. Gonzalez-Garcia for useful discussions. 
NS would like to thank the UK Science and Technology Facilities Council (STFC) for funding this work through support for the Quantum Sensors for the Hidden Sector (QSHS) collaboration under grants ST/T006102/1, ST/T006242/1, ST/T006145/1, ST/T006277/1, ST/T006625/1, ST/T006811/1, ST/T006102/1 and ST/T006099/1. NS is also supported by the National Natural Science Foundation of China (NSFC) Project No. 12047503.  Computations were performed at the Queen's Centre for Advanced Computing, supported in part by the Canada Foundation for Innovation, and at Barkla HPC cluster housed at the University of Liverpool.
JK is supported in part by DOE grant DE-SC0010504.
JB is supported by the Natural Sciences and Engineering
Research Council of Canada (NSERC).
GM acknowledges support from the Arthur B. McDonald Canadian Astroparticle Physics Research Institute and from the UC office of the President via the UCI Chancellor's Advanced Postdoctoral Fellowship. GM also acknowledges partial support from the U.S. National Science Foundation under grant PHY - 2210452 and the National Science and Engineering Research Council of Canada
NR is supported in part by the Natural Sciences and Engineering Research Council (NSERC) of
Canada. 
Research at Perimeter Institute is supported in
part by the Government of Canada through the Department of Innovation, Science and Economic Development Canada and by the Province of Ontario through the Ministry of
Colleges and Universities.
TRIUMF receives federal funding via a contribution agreement with the National Research Council
(NRC) of Canada.

\bibliographystyle{JHEP.bst}

\onecolumngrid

\appendix

\section{Earth Model and Details of Monte Carlo Simulations}
\label{app:MC}

We use the PREM~\cite{dziewonski1981preliminary} model for the density profile of the Earth. The temperature profile is adapted from Ref.~\cite{earle2015physical}.  We also use the Earth composition listed in Table 1 of Ref.~\cite{Bramante:2019fhi}. The isotope abundance and spin information of nuclei relevant for spin-dependent interactions are listed in Table~\ref{tab:nuclearspins}.
\begin{table}[!h]
\centering
\setlength\extrarowheight{2pt}
\setlength{\tabcolsep}{5pt}
\begin{tabular}{l|l|l|l|l}
\hline
\hline
Isotope                                               & Abundance[\%]             & $J$                      & $\langle S_p\rangle$        & $\langle S_n\rangle$ \\  \hline
\Nuc{N}{14}& 99.6& 1& 0.5& 0.5                  \\  \hline
\Nuc{N}{15}& 0.4& 1/2& -0.145& 0.037                \\  \hline
\Nuc{O}{17}& 0.04& 5/2& -0.036& 0.508                \\  \hline
\Nuc{Si}{29}& 4.7& 1/2& 0.016& 0.156                \\  \hline
\Nuc{Al}{27}& 100& 5/2& 0.326& 0.038                \\  \hline
\Nuc{Fe}{57}& 2.12& 1/2& 0      & 0.5                  \\  \hline
\Nuc{Ca}{43}& 0.135& 7/2& 0      & 0.5                  \\  \hline
\Nuc{Na}{23}& 100& 3/2& 0.224& 0.024                \\  \hline
\Nuc{K}{39}& 100& 3/2& -0.196& 0.055                \\  \hline
\Nuc{Mg}{25}& 10& 5/2& 0.040& 0.376                \\  \hline
\Nuc{Ti}{47}& 7.44& 5/2& 0      & 0.21                 \\  \hline
\Nuc{Ti}{49}& 5.44& 7/2& 0      & 0.29                 \\  \hline
\Nuc{Ni}{61}& 1.14& 3/2& 0      & -0.357               \\  \hline
\Nuc{Co}{59}& 100& 7/2& 0.5& 0                    \\  \hline
\Nuc{P}{31}& 100& 1/2& 0.181& 0.032                \\  \hline
\Nuc{S}{33}& 75& 3/2& 0      & -0.3                 \\ \hline
\Nuc{Ge}{73}&7.76&9/2&0.031&0.439\\ \hline
\Nuc{Xe}{129}&26.4&1/2&0.010&0.329\\ \hline
\Nuc{Xe}{131}&21.2&3/2&-0.009&-0.272\\
\hline\hline
\end{tabular}
\caption{The fractional isotope abundance, nuclear spin $J$, average proton spin $\langle S_p\rangle$, and average neutron spin $\langle S_n\rangle$ for isotopes with non-zero spin in the Earth. The average spins for \Nuc{Si}{29}, \Nuc{Al}{27} \Nuc{Na}{23}, \Nuc{Ge}{73}, \Nuc{Xe}{129} and \Nuc{Xe}{131} are taken from Ref.~\cite{Klos:2013rwa}, and the rest from Ref.~\cite{Bednyakov:2004xq} (EOGM $g_A/g_V=1$). The spin information of Ge and Xe isotopes is used to produce the spin-dependent results for CDMSlite and XENON1T.}
\label{tab:nuclearspins}
\end{table}
We take into consideration the effect of the atmosphere on the capture, evaporation and annihilation of dark matter. The atmosphere is assumed to be 100~km thick, beyond which the number density of atmospheric particles is negligibly small. As a consequence, the geometric size of the Earth is $\Ratm=6471$~km. We use the NRLMSISE-00 atmosphere model~\cite{NRLMSISE-00} with the average oxygen, nitrogen, argon, helium mass fractions of 23.18\%, 75.6\%, 1.2\% and $7.25\times 10^{-7}$ respectively.

We assume that dark matter follows a Maxwell-Boltzmann distribution in the rest frame of the Galactic Center, with the velocity dispersion $v_d=270$~km/s, which translates into the rest frame of Earth with the Earth velocity $v_\oplus=220$~km/s. The halo dark matter velocity distribution given by~\cite{Garani:2017jcj}
\begin{equation}
    f(u_\chi)=\sqrt{\dfrac{3}{2\pi}}\dfrac{u_\chi}{v_\oplus v_d}\left(\mathrm{exp}\left[-\frac{3(u_\chi-v_\oplus)^2}{2v_d^2}\right]-\mathrm{exp}\left[-\frac{3(u_\chi+v_\oplus)^2}{2v_d^2}\right]\right)\,.
    \label{eq:halovelocity}
\end{equation}
 A dark matter particle in the halo will be accelerated by the Sun's and Earth's gravitational potential. If an infalling dark matter particle has speed $u_\chi$ when far from the Sun, we assume 
that it will have speed $w(r)=\sqrt{u_\chi^2+v_s^2+v_e(r)^2}$ at a distance $r$ from the Earth center, 
where $v_s=42.2$~km/s is the the escape velocity from the Sun at 1 AU 
and $v_e(r)$ is the escape velocity of the Earth at radius $r$. We neglect the subtleties in the solar velocity boost in the frame of the Earth, see for example~\cite{Griest:1987vc}. 
We have also neglected the effect of the Galactic escape velocity of dark matter at the position of the solar system, 
as the Maxwellian distribution above the escape velocity only makes up a small fraction of dark matter, and the capture of dark matter is dominated by the low velocity part. The geometric capture rate at which all dark matter particles that bombard the Earth are captured is  
\begin{equation}
    C_\oplus^\geom=\pi \Ratm^2\sqrt{\dfrac{8}{3\pi}}\dfrac{\rho_\chi v_d}{m_\chi}\left(1+\dfrac{3v_{\rm esc}^2}{v_d^2}\right)\xi\,,
\end{equation}
where 
\begin{equation}
    \xi=\left[v_d^2e^{-\frac{3v_\oplus^2}{2v_d^2}}+\sqrt{\dfrac{3\pi}{2}}\dfrac{v_d}{v_\oplus}\left(v_{\rm esc}^2+v_\oplus^2+\dfrac{v_d^2}{3}\right)\mathrm{Erf}\left(\sqrt{\dfrac{3}{2}}\dfrac{v_\oplus}{v_d}\right)\right]\left(2v_d^2+3v_{\rm esc}^2\right)^{-1}\,,
\end{equation}  
 We take  $\rho_\chi=0.3$~GeV/cm$^3$ for the local dark matter density. Essentially, the geometric capture rate is the product of the Earth's geometric cross section ($\pi \Ratm^2$), the dark matter flux ($\sim \rho_\chi v_d / m_\chi$) and a factor which accounts for the gravitational version of Sommerfeld-enhancement. We do not consider the indirect capture of particles which first become gravitationally bound to the Sun after scattering in 
the Sun, the Earth, or Jupiter, and are then subsequently captured by the Earth.  Our estimate is, in that sense, conservative.

\section{Comparison with Analytical Approaches for Single Scatter Capture}
\label{app:analytical}
\begin{figure}
    \centering
    \includegraphics[width=0.49\textwidth]{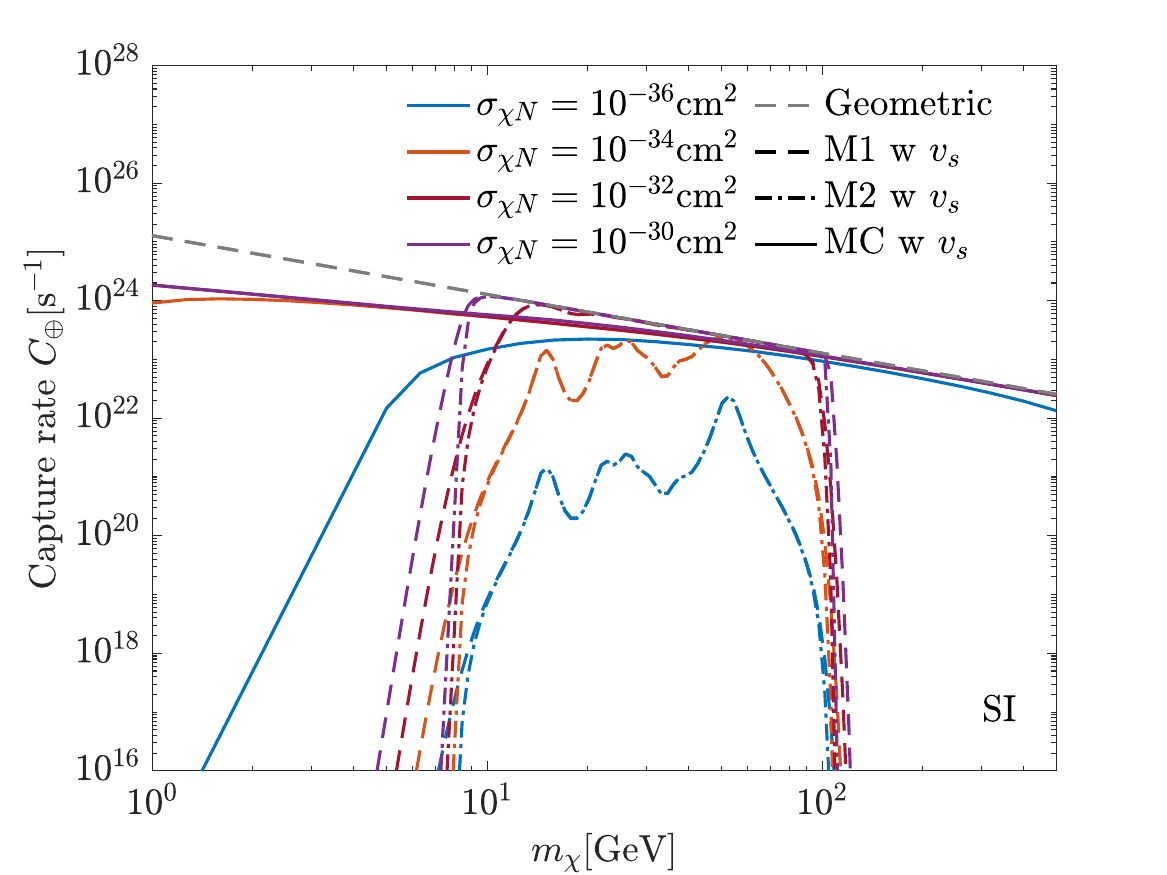}
    \includegraphics[width=0.49\textwidth]{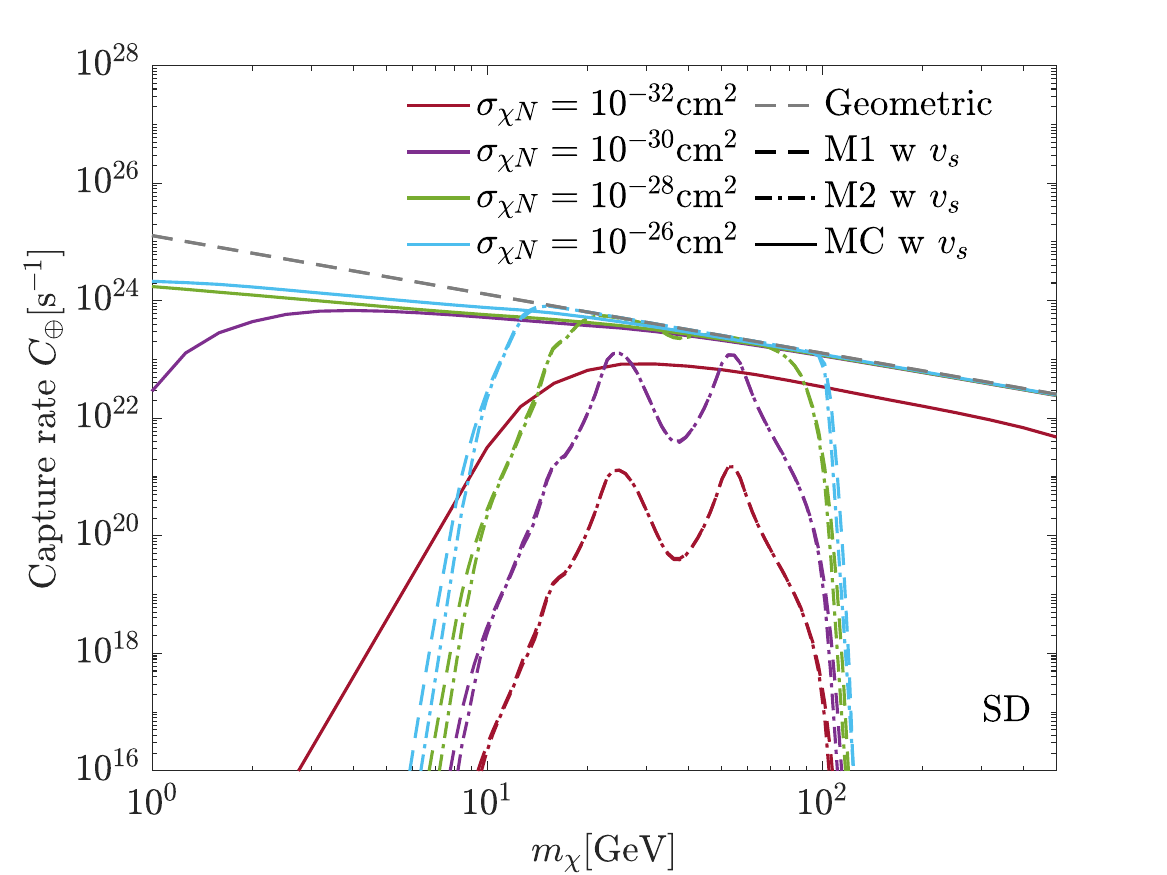} \caption{Total dark matter capture rates in the Earth in the optically thick regime for various dark matter-nucleon scattering cross sections. The dashed, dash-dotted and solid lines show the capture rates from~\cite{Garani:2017jcj} (M1), \cite{Baum:2016oow} (M2) and from Monte Carlo (MC) simulations using \texttt{DaMaSCUS-EarthCapture}, modified from the \texttt{DaMaSCUS} code~\cite{Emken:2017qmp,emken2017damascus}, respectively. Different line colors mark different dark matter-nucleon interaction cross sections. Solar acceleration of halo dark matter is included. The capture fractions from simulations are extrapolated below $10^{-5}$ GeV. See text for details. The dashed gray line depicts the Earth geometric dark matter capture rate when all dark matter particles that encounter the Earth get captured. {\it Left:} Results for spin-independent scattering. {\it Right:} Results for spin-dependent scattering with $a_p=a_n=1$.}
    \label{fig:capanalytical}
\end{figure}

In the analytical approach for single scatter capture, one calculates the probability for a dark matter particle in the halo to scatter once in the Earth and get captured. 
We will review this approach here, and highlight aspects of the capture process which are not fully described in this approach, and are 
more completely revealed by numerical simulation.
For this we primarily follow the approach outlined in~\cite{Garani:2017jcj} (dubbed Model 1, or M1), and briefly mention the approach in~\cite{Baum:2016oow} (dubbed Model 2, or M2). 
If  the interaction between dark matter and 
nuclei is weak, a dark matter particle will likely scatter 
at most once as it crosses the Earth.  
The corresponding capture rate is~\cite{Gould:1987ir}
\begin{equation}
    C_\oplus^{\mathrm{weak}}=\sum\limits_j\dfrac{\rho_\chi}{m_\chi}\int_0^{\RE} 4\pi r^2\int_0^\infty du_\chi f(u_\chi)\dfrac{w(r)}{u_\chi}\int_0^{v_e(r)} R_j^-(w\rightarrow v)|F_j(q)|^2dv\,.
    \label{eq:Cweak}
\end{equation}
The sum runs over all nuclear targets $j$ in the Earth and $R_j^-(w\rightarrow v)$ describes the rate  for a dark matter 
particle of velocity
$w(r)$ to slow down to $v<v_e(r)$ by scattering with nucleus $j$ (see~\cite{Garani:2017jcj} for the explicit expressions of $R_j^\pm$). 
The $R_j^\pm$ are proportional to the dark matter-nucleus scattering cross section.
In the weak scattering limit the capture rate increases with cross section. However, no matter how large the scattering cross section is, the capture rate can never 
exceed $C_\oplus^\geom$, which is the rate at which dark matter is incident on the Earth.
The capture rate can thus be approximated by~\cite{Bernal:2012qh,Garani:2017jcj}
\begin{equation}
    C_\oplus=C_\oplus^{\mathrm{weak}}(1-\exp(-C_\oplus^\geom/C_\oplus^\mathrm{weak}))\,.
    \label{eq:Coplus}
\end{equation}
The capture rates computed from Eq.~\eqref{eq:Cweak} and Eq.~\eqref{eq:Coplus} are shown in Figure~\ref{fig:capanalytical} (dashed lines, M1). The capture is maximized when the dark matter mass matches the mass of the target nucleus, and the typical recoil energy is maximized, resulting in the peaks around 10-100~GeV. For spin-independent scattering, from left to right we can identify the the peaks caused by scattering with O, with Mg, and 
with Si and Fe. For spin-dependent scattering, there are instead two peaks arising from scattering with \Nuc{Si}{29}, \Nuc{Al}{27}, \Nuc{Mg}{25}, as well as \Nuc{Fe}{57}.  At large cross sections, the peaks are smeared out as the capture saturates the geometric rate.

For comparison we also show in Figure~\ref{fig:capanalytical} the results extrapolated from~\cite{Baum:2016oow} (dash-dotted lines, M2), where the capture rate is computed using Eq.~\eqref{eq:Coplus}, but with $C_\oplus^{\mathrm{weak}}$ given by~\cite{Baum:2016oow}
\begin{equation}
    C_\oplus^{\mathrm{weak}}=\sum\limits_j\dfrac{\rho_\chi}{m_\chi}\sigma_{j,0}\int_0^{\RE} 4\pi r^2 v_e^2(r)n_j(r)\int_0^{u_{\chi,\max}} du_\chi \dfrac{f(u_\chi)}{u_\chi}\left(1-\dfrac{u_\chi^2+v_s^2}{u_{\chi,\max}^2+v_s^2}\right)\,,
    \label{eq:CweakBaum}
\end{equation}
with $n_j$ the number density of an isotope $j$ in the Earth. Note that we have included the solar gravitational acceleration explicitly but discarded the momentum transfer dependence of the form factors. In a single scattering, the maximum momentum transfer is $q_{\max}=2\mu_{A_j}w$, causing the maximum kinetic energy loss $E_{R,\max}=2\mu_{A_j}^2w^2/m_{A_j}$. 
If the dark matter particle is captured, the dark matter kinetic energy after scatter must satisfy the relation $E_{\chi,f}=\frac{1}{2}m_\chi w^2-E_{R,\max}\leq \frac{1}{2}m_\chi v_e^2(r)$.  
This caps the dark matter initial velocity at
\begin{equation}
    w\leq \left\lvert \dfrac{m_\chi+m_{A_j}}{m_\chi-m_{A_j}}\right\rvert v_e(r)\,.
\end{equation}
With the relation $w(r)=\sqrt{u_\chi^2+v_s^2+v_e(r)^2}$, this translates to the maximum dark matter halo speed such that 
single-scatter capture is possible:
\begin{equation}
    u_{\chi,\max}=\sqrt{\dfrac{4m_\chi m_{A_j}}{(m_\chi-m_{A_j})^2}v_e^2(r)-v_s^2}\,.
    \label{eq:uchimax}
\end{equation}
If the dark matter speed (at infinity) is greater than $u_{\chi,\max}$ in the halo, its speed after in-fall to radius 
$r$ will be such that, even if scatter results in the maximum momentum transfer, the dark matter still will not have dropped below the escape velocity.
Hence, gravitational capture with a single scatter is highly improbable.
This is intuitive as the dark matter velocity change in scattering is proportional to the momentum transfer $q$, which decreases with smaller dark matter mass. 
It is also understood that the square root in Eq.~\eqref{eq:uchimax} has to be 
real,
which determines the minimum captured dark matter mass in single scattering
\begin{equation}
    m_{\chi,\min}\simeq m_{A,\min}\left(1-2\dfrac{v_e}{v_s}+2\dfrac{v_e^2}{v_s^2}-\dfrac{v_e^3}{v_s^3}\right)\simeq 0.59~m_{A,\min}\,,
    \label{eq:mchimin}
\end{equation}
where we have dropped higher powers of $v_e/v_s$. In the second equation we have used $v_e=11.2$~km/s, the escape speed at the surface of the Earth.  
If scattering occurred at the center of the Earth, we would instead find $m_{\chi, \min} \sim 0.5~m_{A,\min}$. 
Using $\Nuc{N}{14}$ puts the minimum dark matter mass at 7.7~GeV, while He might facilitate the capture of dark matter as light as 2.2~GeV, despite the fact that the He abundance in the atmosphere is extremely small.
Similarly, there is a maximum dark matter mass such that capture is possible with a single scatter, given by
\begin{equation}
    m_{\chi,\max}\simeq m_{A,\max}\left(1+2\dfrac{v_e}{v_s}+2\dfrac{v_e^2}{v_s^2}+\dfrac{v_e^3}{v_s^3}\right)\simeq 1.64~ m_{A,\max}\, .
    \label{eq:mchimin}
\end{equation}

In Figure~\ref{fig:capanalytical} we also compare the analytical capture rates obtained from Eq.~\eqref{eq:Coplus} with Monte Carlo (MC) simulations. The MC capture rates are computed with $C^{\rm MC}_\oplus=f_CC_\oplus^{\rm geom}$. Due to the limitation of the number of dark matter samples in the simulations, we do not have reliable data below $f_C=10^{-5}$. We therefore extrapolate the capture fraction to low dark matter masses and compute the corresponding capture fraction for $\sigma_0\leq 10^{-32}$~cm$^2$. The extrapolation does not affect our dark matter bounds since we are primarily interested in $m_\chi\gtrsim$~GeV and large cross sections, where the capture fraction is always larger than $10^{-5}$. We find that using Eq.~\eqref{eq:Coplus} to extrapolate the capture rate beyond the weak scattering limit tends to underestimate the capture rate, unless this approximation reaches the 
saturation limit.

The main difference between the M1 and M2 analyses is that the M1 analysis incorporates the thermal motion of nuclei in the scattering process.
In both formalisms, the weak capture rate is proportional to the scattering cross section.  
Although the M2 analysis does not account for the thermal motion of Standard Model nuclei, it agrees with M1 remarkably well at all dark matter masses once the 
effects of the solar gravitational potential are included. 
It is important to note that the upper and lower limits on the dark matter mass at which single-scatter capture is possible only appear because the Earth-DM system is in the 
external gravitational potential of the Sun.  In the absence of this external gravitational potential, single-scatter capture would be possible for any dark matter mass, for some 
choice of the dark matter speed far from the Earth.  The Sun's gravitational potential has a major impact on the behavior of the capture rate at large scattering cross section.  Given the approximation in Eq.~\eqref{eq:Coplus}, 
we see that if single-scatter capture of any non-zero fraction of the incident dark matter flux is kinematically possible, then at sufficiently large scattering cross section, 
the capture rate will saturate to the geometric capture rate.  But if the dark matter mass is such that single-scatter capture is kinematically impossible, then the capture rate in 
Eq.~\eqref{eq:Coplus} will vanish, no matter how strong the coupling is.

\section{Effects of solar gravitational potential}
\label{app:solargravity}

\begin{figure}
    \centering
    \includegraphics[width=0.48\textwidth]{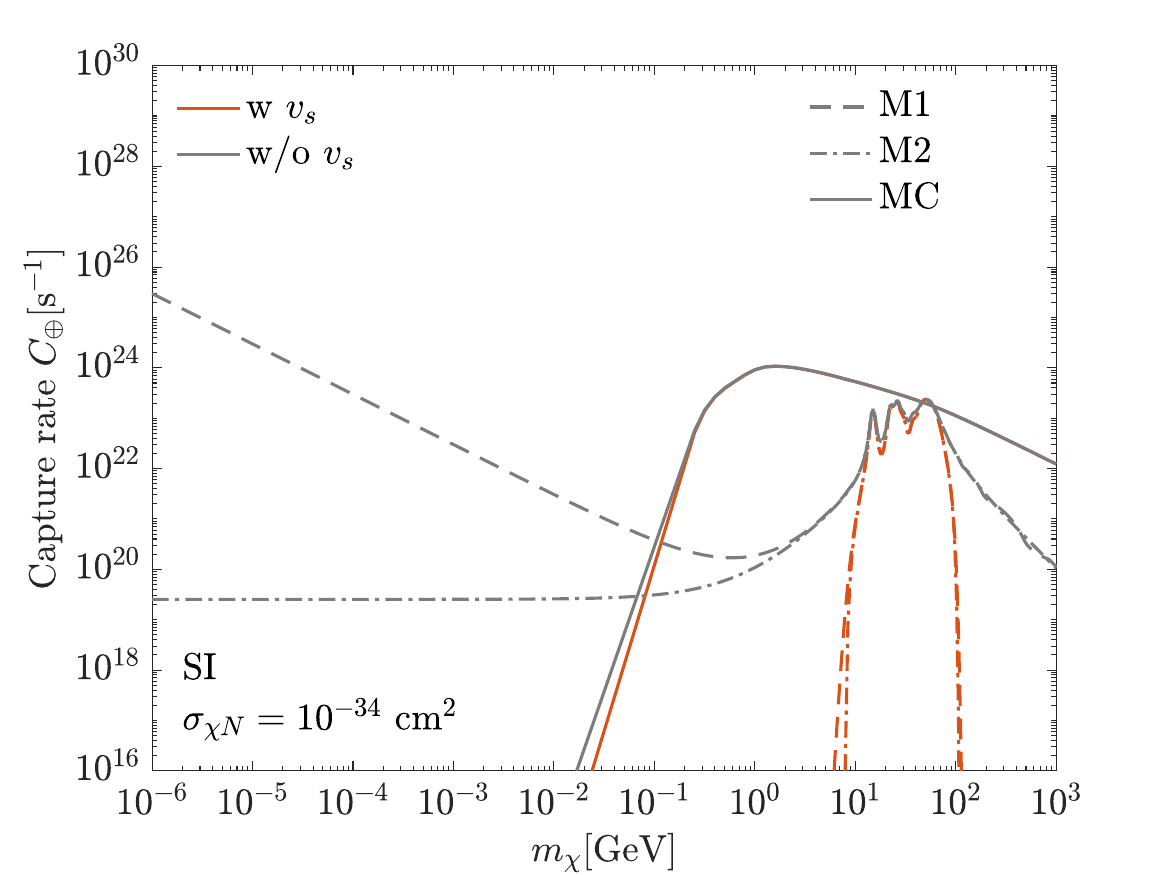}
    \includegraphics[width=0.48\textwidth]{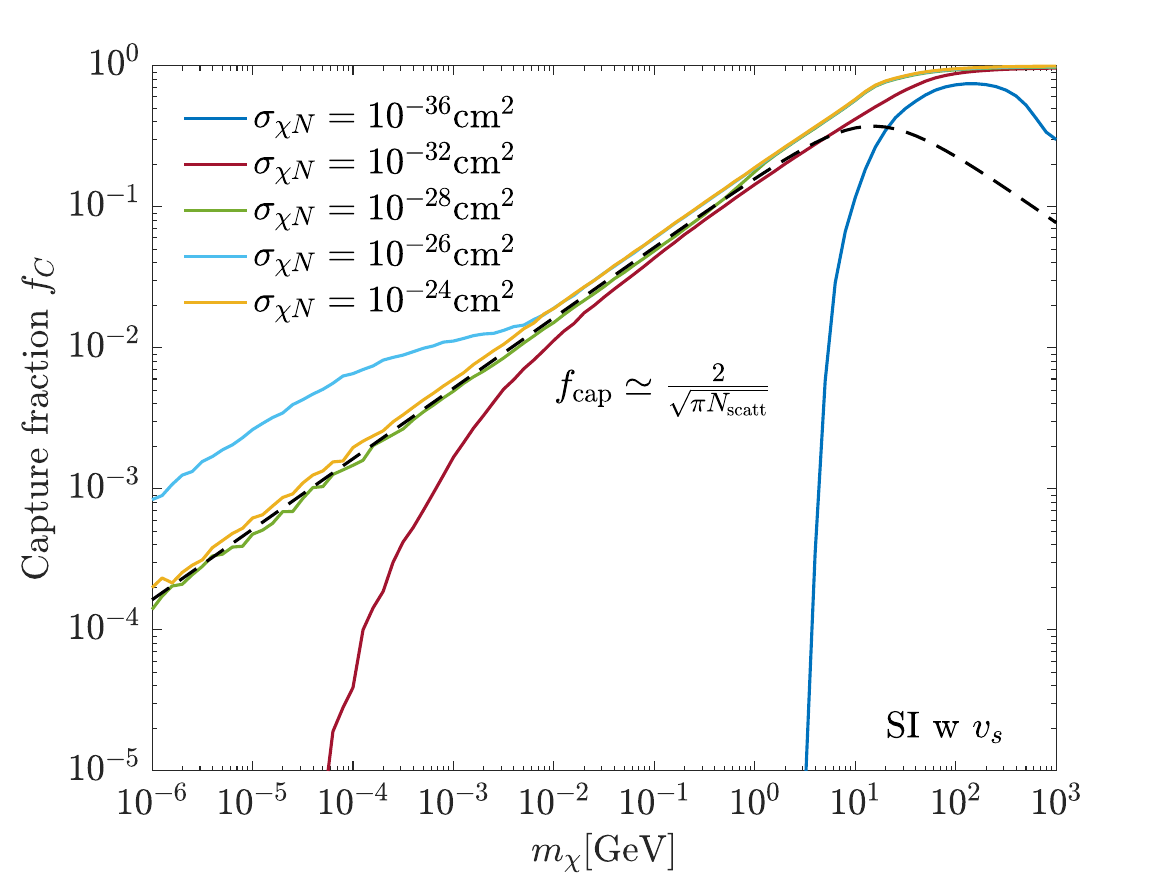}    
    \caption{Total dark matter capture rates in the Earth in the optically thick regime assuming spin-independent interactions. {\it Left:} The dashed, dash-dotted and solid lines show the capture rates from~\cite{Garani:2017jcj} (M1), \cite{Baum:2016oow} (M2) and from Monte Carlo simulations using \texttt{DaMaSCUS-EarthCapture}, respectively. We assume the spin-independent cross section $\sigma_{\chi N}^{\rm SI}= 10^{-34}$~cm$^2$. The red curves include the effect of solar gravitational acceleration, while the grey curves not. See text for details. {\it Right:} The fraction of dark matter particles that are captured among those impinge on the Earth as a function of dark matter mass. Various cross sections are depicted by lines of different colors. The dashed black line corresponds to reflection factor in Eq.~\eqref{eq:fcap} assuming oxygen target.}
    \label{fig:comparevs}
\end{figure}

For completeness, we also study the effect of solar gravitational potential on the capture process, and display in Figure~\ref{fig:comparevs} left panel the capture rates with and without including solar acceleration. The latter is simply achieved by setting $v_s=0$ in Eqs.~\eqref{eq:Cweak} and~\eqref{eq:CweakBaum}. The agreement between M1 and M2 analytical results remain robust in the high mass limit above 10~GeV. However, in the low mass end, the capture rate in M2 tends to be flat, while the M1 rate scales as $1/m_\chi$ when solar acceleration is ignored. 
In stark contrast to the analytical results, the capture rates in MC simulations change marginally with or without including $v_s$. As we will show in the next section, the number of collisions of dark matter before capture in the optically thick limit is only logarithmically sensitive to the dark matter initial velocity. 
As the capture probability of low mass dark matter is related to the number of collisions, it is not surprising that increasing the 
lowest dark matter velocity has little effect on the results.

Interestingly, we also find the analytical results yield a result closer to the Monte Carlo result at high masses if one fails to take into account the solar gravitational potential than 
if one correctly accounts for it.  If one does not account for the solar gravitational potential, then single-scatter capture would be possible, which is extrapolated to a capture rate that saturates the geometric rate at sufficiently large cross section.  However, the absence of the solar gravitational potential implies that multi-scatter capture is not correctly accounted for.
Note that the capture rate is not correctly modeled even at large mass unless it has saturated 
the geometric rate.
On the other hand, at low masses, the extrapolation in Eq.~\eqref{eq:Coplus} 
does not match the Monte Carlo result regardless of whether or not solar acceleration is included, because it does not account for the reflection of dark matter from the Earth.  
Note, however, that such low-mass dark matter will in any case not contribute significantly to 
Earth heating, because of evaporation.

\section{Comparison with Analytical Approaches for Multi Scatter Capture}
\label{app:multianalytical}

The analytical formalism for multi-scatter dark matter capture has been established in Refs.~\cite{Leane:2022hkk,Kouvaris:2010vv,Bramante:2017xlb,Dasgupta:2019juq,Ilie:2020vec}, which we call Model 3 or M3. We revisit the formalism in this work with some novel treatment in both analytical and numerical aspects. The capture rate after scattering $N$ times in the Earth is
\begin{equation}
    C_N=f_{\rm cap}\pi R^2p_N(\tau)\dfrac{\sqrt{6}\rho_\chi}{3\sqrt{\pi}m_\chi v_\chi}\left[(2v_\chi^2+3v_{\rm esc}^2)-(2v_\chi^2+3v_N^2)\exp\left(-\dfrac{3(v_N^2-v_{\rm esc}^2)}{2v_\chi^2}\right)\right]\,,
    \label{eq:CN}
\end{equation}
where $v_\chi$ is the average halo dark matter velocity taken to be 270~km/s, and $v_{\rm esc}=11.2$~km/s is the escape velocity at Earth. $v_N=v_e(1-\langle z\rangle\beta)^{-N/2}$ is the maximum velocity for dark matter arriving at the Earth which could be captured after $N$ times (note that we have factor of 2 difference from~\cite{Leane:2022hkk}). We cut $v_N$ at 800~km/s inspired by the local galactic escape velocity of dark matter, regardless of $N$. It is found that $\langle z\rangle\simeq 0.5$ for isotropic scattering and $\beta\equiv 4m_\chi m_A/(m_\chi+m_A)$, where $m_A$ is the nuclear target in the scattering. The probability of capture after $N$ scattering
\begin{equation}
    p_N(\tau)=2\int_0^1 dy\dfrac{ye^{-y\tau}(y\tau)^N}{N!}\,,
    \label{eq:pN}
\end{equation}
where the optical depth
\begin{equation}
    \tau=\dfrac{3}{2}\dfrac{\sigma}{\sigma_{\rm sat}}\,.
    \label{eq:tau}
\end{equation}
The scattering cross section $\sigma$ with the target of mass number $A$ is given in the main text for spin-independent and spin-dependent interactions. $\sigma_{\rm sat}=\pi R^2/N_t$ is the saturation cross section with $N_t$ the number of scattering targets (note that the definition of $N_t$ is different from~\cite{Leane:2022hkk}). As only one scattering target is dealt with in Eq.~\eqref{eq:CN}, we assume the Earth is made of entirely oxygen or iron so that $N_t=M_\oplus/m_A$ for spin-independent scattering, and $M_\oplus$ is the mass of the Earth. In this scenario, at high scattering cross section, dark matter scatters in the Earth atmosphere or crust where the scattering against oxygen or nitrogen dominates. For lower cross section, dark matter may be stopped in the mantle or the core, where the scattering with iron  contributes the most due to the large mass number (cross section). If the dark matter mass is smaller or comparable to the target mass, dark matter is likely to be reflected after multiple scattering and leave the Earth. This is accounted for in~\cite{Leane:2022hkk} by including the reflection factor
\begin{equation}
    f_{\rm cap}\simeq \dfrac{2}{\sqrt{\pi N_{\rm scatt}}}=\left[\dfrac{2}{\pi}\log(1-\langle z\rangle \beta)/\log\left(\dfrac{v_{\rm esc}}{v_\chi}\right)\right]^{1/2}\,,
    \label{eq:fcap}
\end{equation}
which is used to compute the capture rate when $m_\chi<m_A$. When the dark matter mass is comparable to the target mass, the kinematics could be complicated and simulations are required, as is done in~\cite{Leane:2022hkk}. For simplicity, we set $f_{\rm cap}=1$ for $m_\chi\geq m_A$.  The total capture rate by summing over all possible number of scattering is
\begin{equation}
    C_\oplus^{\rm multi}=\sum\limits_{N=1}^\infty C_N\,.
    \label{eq:Csum}
\end{equation}
The capture probability in Eq.~\eqref{eq:pN} peaks at $N\sim \tau$, and the sum in Eq.~\eqref{eq:Csum} can generally be truncated at $N_{\max}\simeq 2\tau$, beyond which $p_N\rightarrow 0$. However, if $\tau$ is large, computing $p_N$ at large $N$ is numerically difficult. Eq.~\eqref{eq:CN} can be greatly simplified in the limit $v_{\rm esc}\gg v_\chi$ and $m_\chi\gg m_A$ as in~\cite{Bramante:2017xlb}, which unfortunately do not hold for the Earth. Nevertheless, the integral can still be carried out precisely if we notice the integrand in Eq.~\eqref{eq:CN} is close to a Poisson distribution, which is closely resembled by the normal distribution at $y\tau\gg 1$. We can therefore replace the integrand by a Gaussian function and integrate from $y\tau=x_{\min}$ to $\tau$, which yields
\begin{equation}
\begin{split}
    p_N(\tau)\simeq &\dfrac{N+1}{\tau^2}\left[\mathrm{Erf}\left(\dfrac{N+1-x_{\min}}{\sqrt{2x_{\min}}}\right)-\mathrm{Erf}\left(\dfrac{N+1-\tau}{\sqrt{2\tau}}\right)\right.\\
    &\left.-e^{2(N+1)}\left(\mathrm{Erf}\left(\dfrac{N+1+x_{\min}}{\sqrt{2x_{\min}}}\right)-\mathrm{Erf}\left(\dfrac{N+1+\tau}{\sqrt{2\tau}}\right)\right)\right]\,.
\end{split}
\label{eq:pNapprox}
\end{equation}
At $y\tau\gg 1$, the contribution to the integral from $0\leq y\tau \leq x_{\min}$ is small as long as $x_{\min}\ll 1$. We choose $x_{\min}=10^{-3}$, which precisely reproduces the full integral if $\tau>50$. Conservatively, we numerically calculate $p_N$ for $\tau<100$, and use Eq.~\eqref{eq:pNapprox} for larger $\tau$.

 For spin-dependent scattering, the largest contribution to the capture comes from scattering with \Nuc{Si}{29}, \Nuc{Al}{27}, \Nuc{Mg}{25} if dark matter is mainly stopped in the Earth crust or mantle, and \Nuc{Fe}{57} if dark matter matter is stopped in the core. At very large cross section $\sigma^{\rm SD}_{\chi N}\gg 10^{-24}$~cm$^2$, the captured dark matter mainly scatters with $\Nuc{N}{14}$ in the atmosphere. We mainly explore the first two scenarios. In the former case, \Nuc{Si}{29}, \Nuc{Al}{27}, \Nuc{Mg}{25} have similar nuclear mass (hence kinematics) and nuclear response. We therefore set $N_t=N_{\rm \Nuc{Si}{29}}+N_{\rm \Nuc{Al}{27}}+N_{\rm \Nuc{Mg}{25}}$, and use the angular momentum and average spin of \Nuc{Al}{27} to compute the nuclear scattering cross section. For the latter we set $N_t=N_{\rm \Nuc{Fe}{57}}$ and compute the cross section correspondingly.

\begin{figure}
    \centering
    \includegraphics[width=0.48\textwidth]{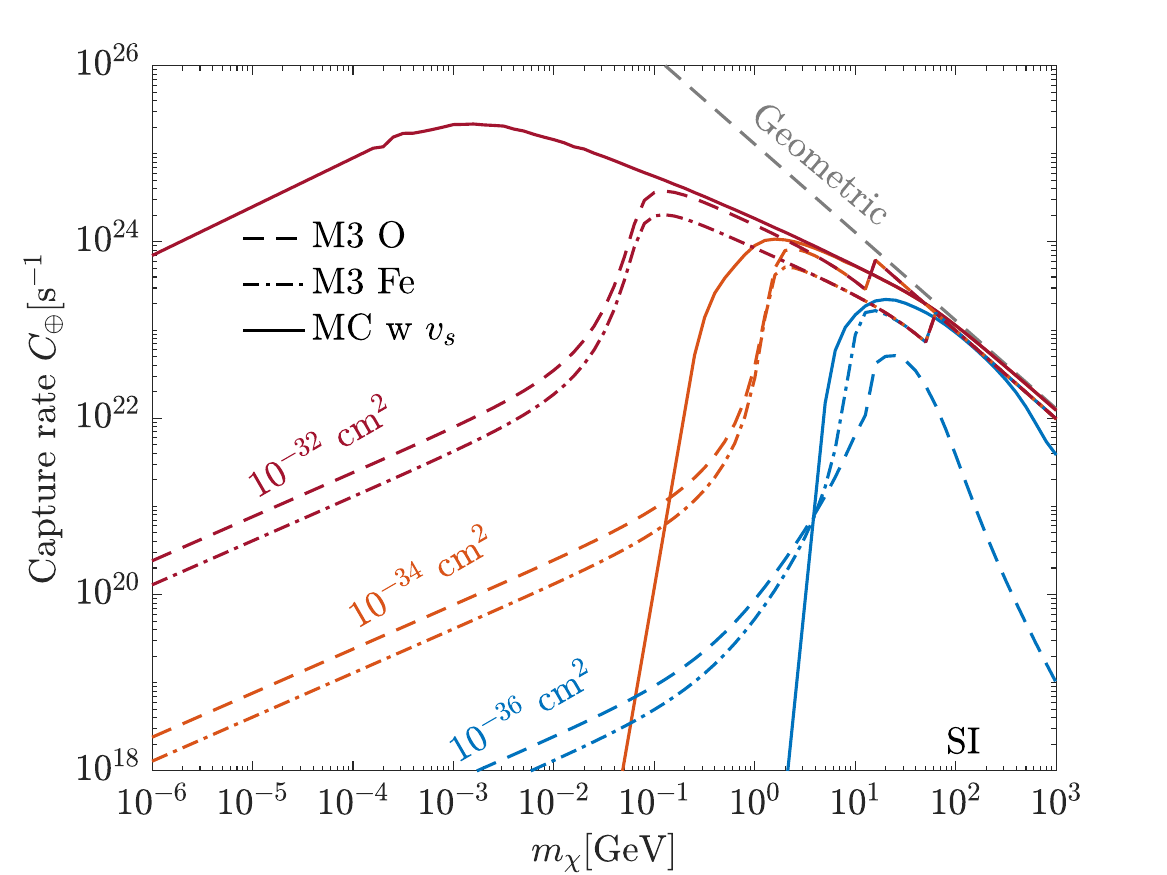}
    \includegraphics[width=0.48\textwidth]{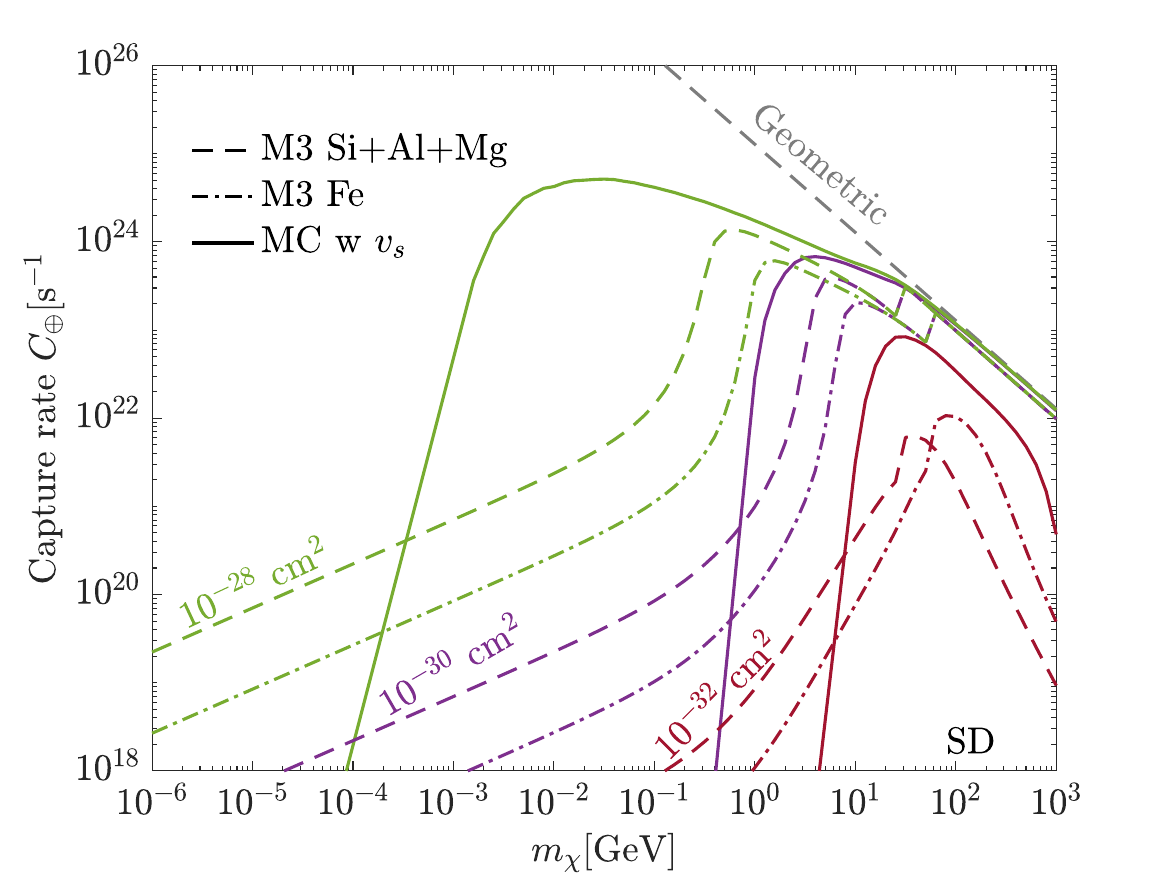}    
    \caption{Total dark matter capture rates in the Earth in the optically thick regime for various dark matter-nucleon scattering cross sections. The solid lines show the capture rates from Monte Carlo (MC) simulations using \texttt{DaMaSCUS-EarthCapture}, and broken lines depict the results from multi-scatter capture formalism, implemented and discussed in this work (dubbed M3).  {\it Left:} Results for spin-independent (SI) scattering. The dashed line assumes the Earth is made of O, and the dash-dotted lines assumes Fe instead. {\it Right:} Results for spin-dependent (SD) scattering with $a_p=a_n=1$. The dashed lines account for the SD scattering against \Nuc{Si}{29}, \Nuc{Al}{27} and \Nuc{Mg}{25}, and the dash-dotted lines considers dark matter scattering with \Nuc{Fe}{57}. See text for details.}
    \label{fig:Cmulti}
\end{figure}

The comparison between Monte Carlo simulations and the multi-scatter model is presented in Figure~\ref{fig:Cmulti}. The kinks in the model predictions come from the abrupt shutoff of $f_{\rm cap}$
at $m_\chi=m_A$. At low cross section ($\sigma_{\chi N}^{\rm SI}\lesssim 10^{-36}$~cm$^2$ and $\sigma_{\chi N}^{\rm SD}\lesssim 10^{-32}$~cm$^2$), the analytical formalism underestimates the capture rate at high dark matter mass and overestimates the rate at low mass. As cross section increases, the model reproduces the MC results well at high masses, regardless of the choice of chemical compositions in the Earth. However, for low dark matter mass significant discrepancy still remains before the reflection factor in Eq.~\eqref{eq:fcap} is saturated. As can be seen from Figure~\ref{fig:comparevs}, the reflection factor alone correctly predicts the capture fraction at very high cross section $\sigma^{\rm SI}_{\chi N}\gg 10^{-28}$~cm$^2$ for $m_\chi\ll m_A$, except for the resonant capture behavior that will be discussed.

Although only one scattering target is considered in Eq.~\eqref{eq:CN}, analytical formalism for multiple-target capture is investigated in~\cite{Ilie:2021iyh}. As assumptions similar to the multi-scatter model above were made in computing the capture rates, we do not expect a substantial improvement when comparing with the Monte Carlo results. A dedicated comparison using multiple-target capture is left for future work.

\section{Resonant Light Dark Matter Capture}
\label{app:resonantcapture}
\begin{figure}
    \centering
    \includegraphics[width=0.48\textwidth]{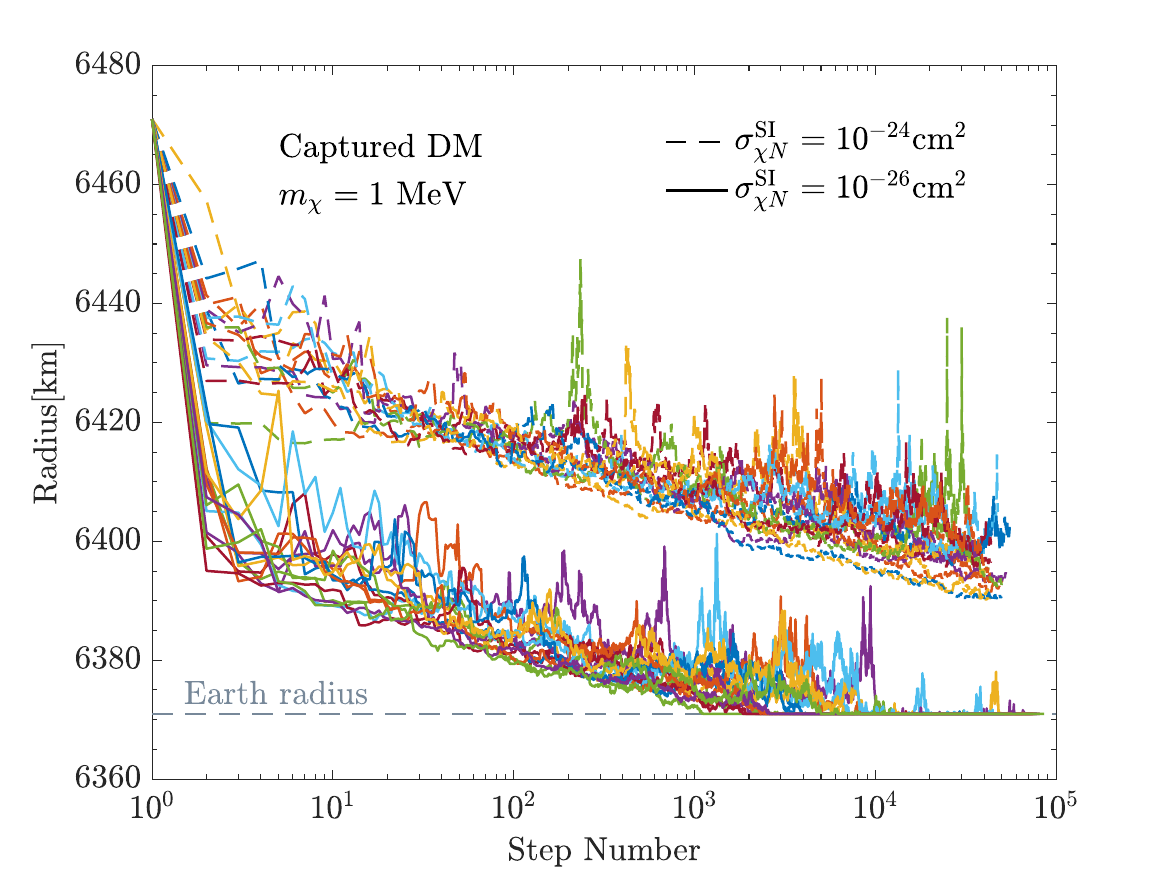}
    \includegraphics[width=0.48\textwidth]{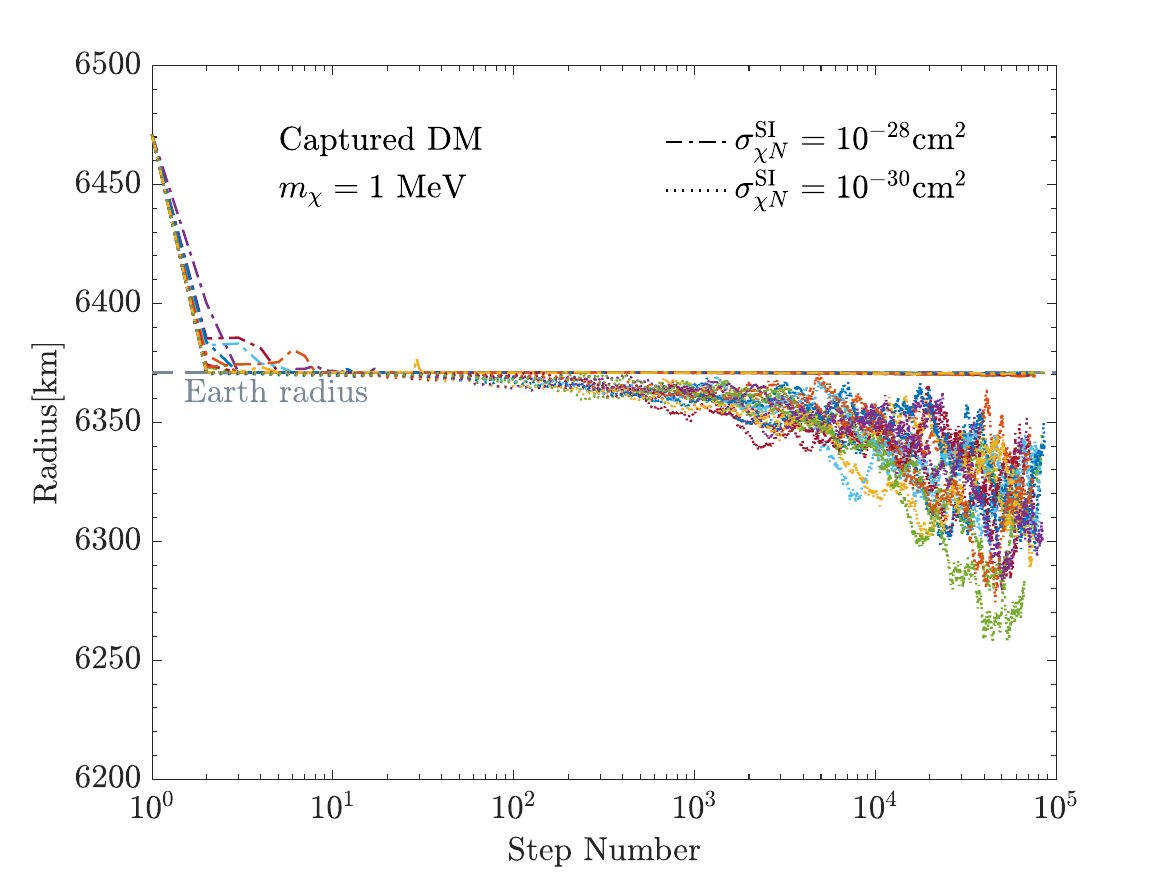}    
    \caption{Example trajectories of MeV mass dark matter captured inside the Earth or the Earth's atmosphere from~\texttt{DaMaSCUS-EarthCapture} simulations, for various per-nucleon spin-independent scattering cross sections.  Dark matter starts from the top of the atmosphere with $r=6471$~km. At each step, dark matter scatters with a Standard Model nucleus at a specific radius and finally stops at a radius $r<6471$~km.}
    \label{fig:trajcap}
\end{figure}

\begin{figure}
    \centering
    \includegraphics[width=0.48\textwidth]{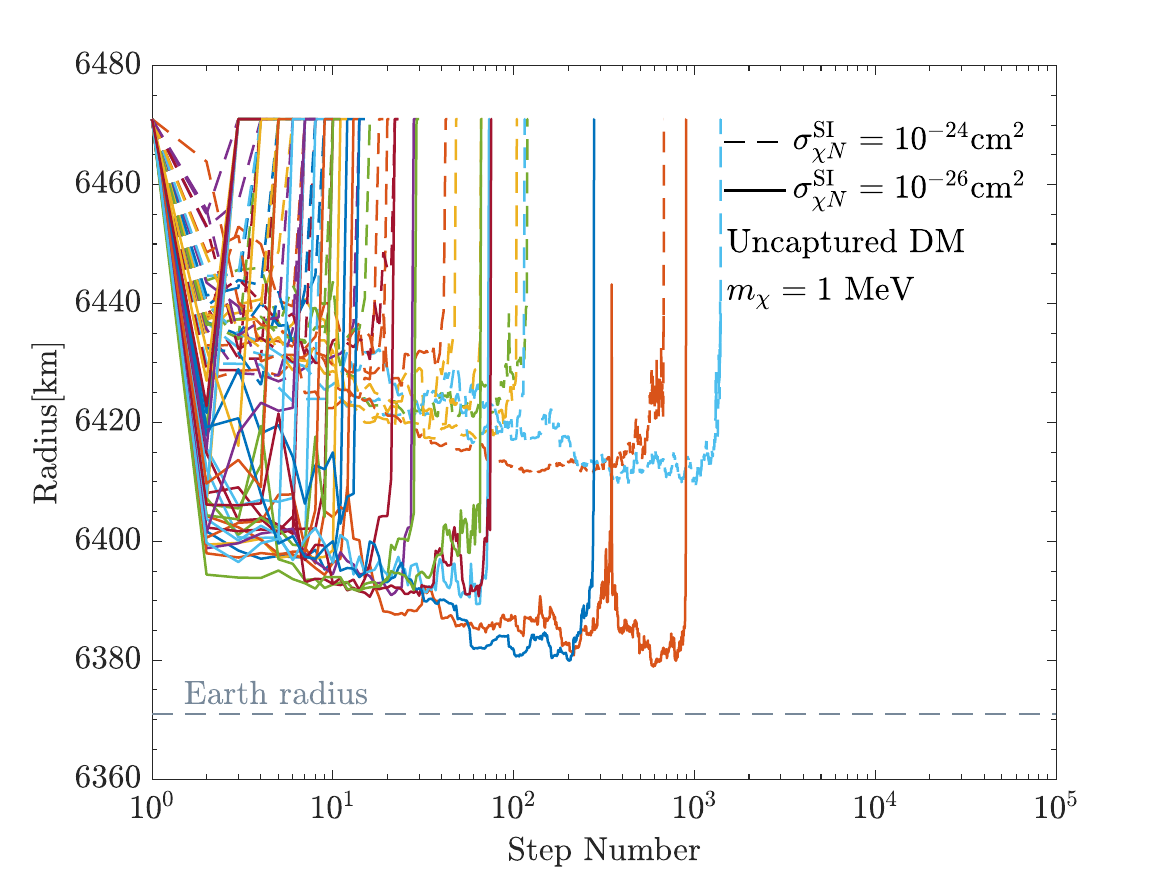}
    \includegraphics[width=0.48\textwidth]{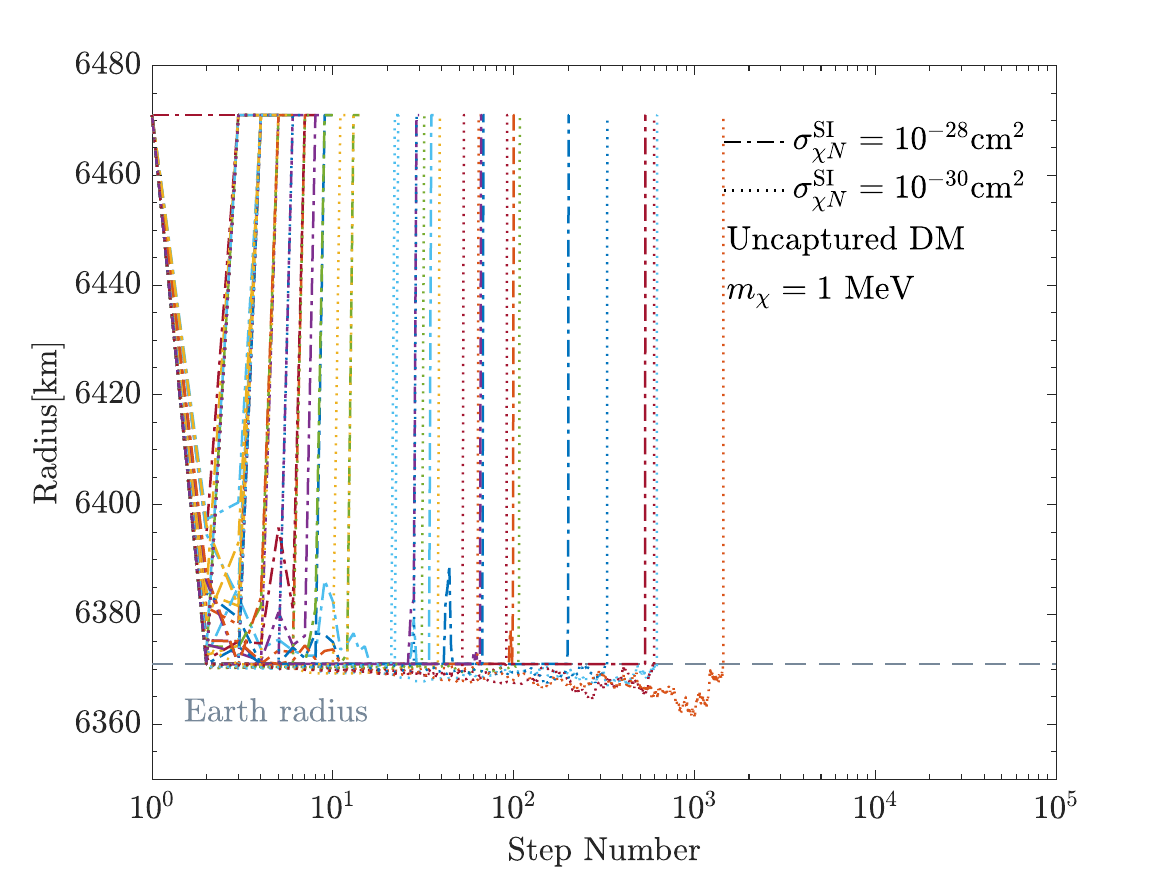}    
    \caption{Example trajectories of MeV mass dark matter {\it NOT} captured inside the Earth or the Earth's atmosphere from~\texttt{DaMaSCUS-EarthCapture} simulations,  for various per-nucleon spin-independent scattering cross sections. Dark matter starts from the top of the atmosphere with $r=6471$~km. At each step, dark matter scatters with Standard Model nucleus at a specific radius and finally exits at a radius $r=6471$~km.}
    \label{fig:trajuncap}
\end{figure}

As shown in the main text, for spin-independent scattering the capture fraction is peaked near $10^{-26}$~cm$^2$ for dark matter mass $m_\chi\lesssim 1$~GeV. To understand this behavior, we show the trajectories of dark matter captured in the Earth or Earth's atmosphere in Figure~\ref{fig:trajcap} and Figure~\ref{fig:trajuncap}. Assuming dark matter with a velocity $w$ scatters with a nucleus at rest, the momentum transfer $q\sim \mu_{\chi A}w\sim m_\chi w$ for light dark matter $m_\chi\ll m_A$. The dark matter kinetic energy loss in each scattering is just the recoil energy of the nucleus $E_R\sim q^2/(2m_A)$. Therefore, the speed change of dark matter in a typical scattering is
\begin{equation}
    \dfrac{\Delta w^2}{w^2}\sim -\dfrac{m_\chi}{m_A}\,,
\end{equation}
\ie~the speed change is smaller for lighter dark matter, where more scattering is necessary to bring the dark matter velocity down below the escape velocity. The number of scatters required in the capture process is therefore
\begin{equation}
    N_{\rm scatt}\simeq 2\dfrac{m_A}{m_\chi}\ln\dfrac{w_i}{w_f}\,.
\end{equation}
If dark matter scatters with nitrogen in the atmosphere, with $w_i\sim 200$~km/s and $w_f\sim 10$~km/s, then we find $N_{\rm scatt}\sim 8\times 10^4$, consistent with the number of steps in Figure~\ref{fig:trajcap}. The fraction of dark matter that is captured in the Earth after scattering and reflection can be estimated as~\cite{Leane:2022hkk,Neufeld:2018slx}
\begin{equation}
    f_C\simeq \dfrac{2}{\sqrt{\pi N_{\rm scatt}}}\simeq \sqrt{\dfrac{4}{\pi}\dfrac{m_\chi}{m_A}\left(\ln\left(1+\dfrac{v_d^2+v_s^2}{v_{\rm esc}^2}\right)\right)^{-1}}\,.
    \label{eq:fCestimate}
\end{equation}
We see that $f_C\sim f_{\rm cap}$ up to a small numerical factor for $m_\chi\ll m_A$, which also reproduces the high cross section saturation capture fraction. However, Eq.~\eqref{eq:fCestimate} also deviates significantly from the capture fraction of $\sigma_{\chi N}\sim 10^{-26}$~cm$^2$. We will explain this behaviour below.

The dark matter velocity after one scattering is
\begin{equation}
    \vec{w}'=\dfrac{m_Aw\uvec{n}+m_\chi \vec{w}}{m_\chi+m_A}\,,
\end{equation}
where $\uvec{n}$ is the direction vector of dark matter in the center of mass frame. Defining $\cos\alpha=\uvec{n}\cdot\uvec{w}$, $\cos\alpha$ is uniformly distributed 
between -1 and 1 in the absence of the form factor. In the lab frame, we have the scattering angle
\begin{equation}
    \cos\alpha'=\dfrac{m_A\cos\alpha+m_\chi}{\sqrt{m_A^2+m_\chi^2+2m_Am_\chi\cos\alpha}}\,.
\end{equation}
In the light dark matter limit $m_\chi\ll m_A$, $\cos\alpha'\simeq \cos\alpha$, and dark matter is deflected randomly between 0 and $\pi$. As we can see from Figure~\ref{fig:trajcap} and Figure~\ref{fig:trajuncap}, light dark matter scatters multiple times before captured. Amid these scatterings, for large enough cross section $\sigma_{\chi N}\sim 10^{-24}$~cm$^2$, light dark matter is likely to be reflected in the Earth's atmosphere and escape, while for small enough cross section $\sigma_{\chi N}\lesssim 10^{-28}$~cm$^2$, dark matter is reflected in the Earth's crust or mantle before they leave. For cross sections in between, dark matter reflected in the Earth's crust might be deflected back in the atmosphere, causing dark matter to be finally captured after bouncing back and force. The capture probability is therefore the sum of all possible trajectories (or paths) along which dark matter is captured. The path integral is apparently larger for intermediate cross section, where dark matter's trajectories cross different media, than for very high or very low cross section, where the trajectories mostly cross one medium. The regime where path integral maximizes corresponds to the peak around $10^{-26}$~cm$^2$.

It is also worth noting that more intense scattering is demanded for smaller dark matter mass, and the peak  moves to higher cross section. The peak is also more pronounced for lighter dark matter, as more scatterings before capture facilitate more viable trajectories, which enhances the overall path integral compared with other cross sections. The peak is also missing for spin-dependent interactions, where the cross section up to $10^{-24}$~cm$^2$ is not high enough to raise the peak.  

\section{Capture fractions for different types of interaction}
\label{app:variousinteractions}
\begin{figure}[!htb]
    \centering
    \includegraphics[width=0.48\textwidth]{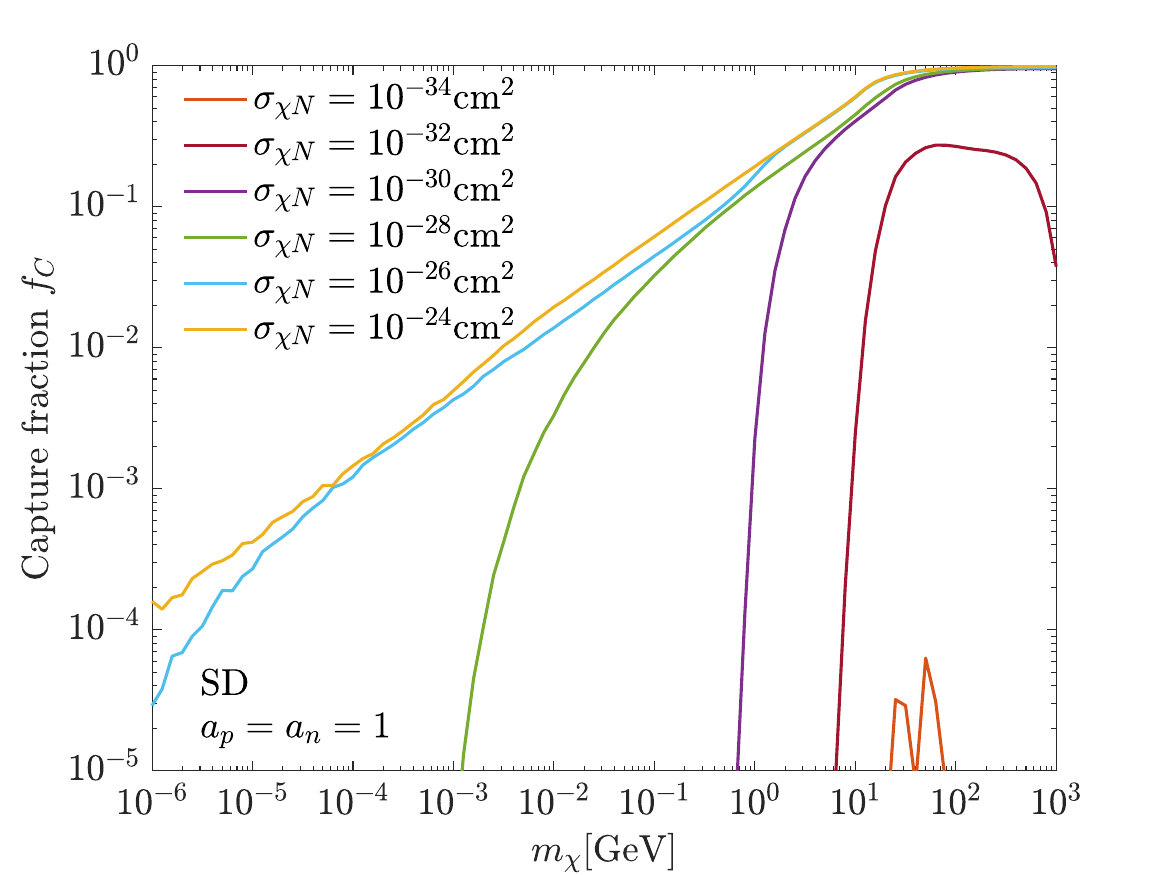}
    \includegraphics[width=0.48\textwidth]{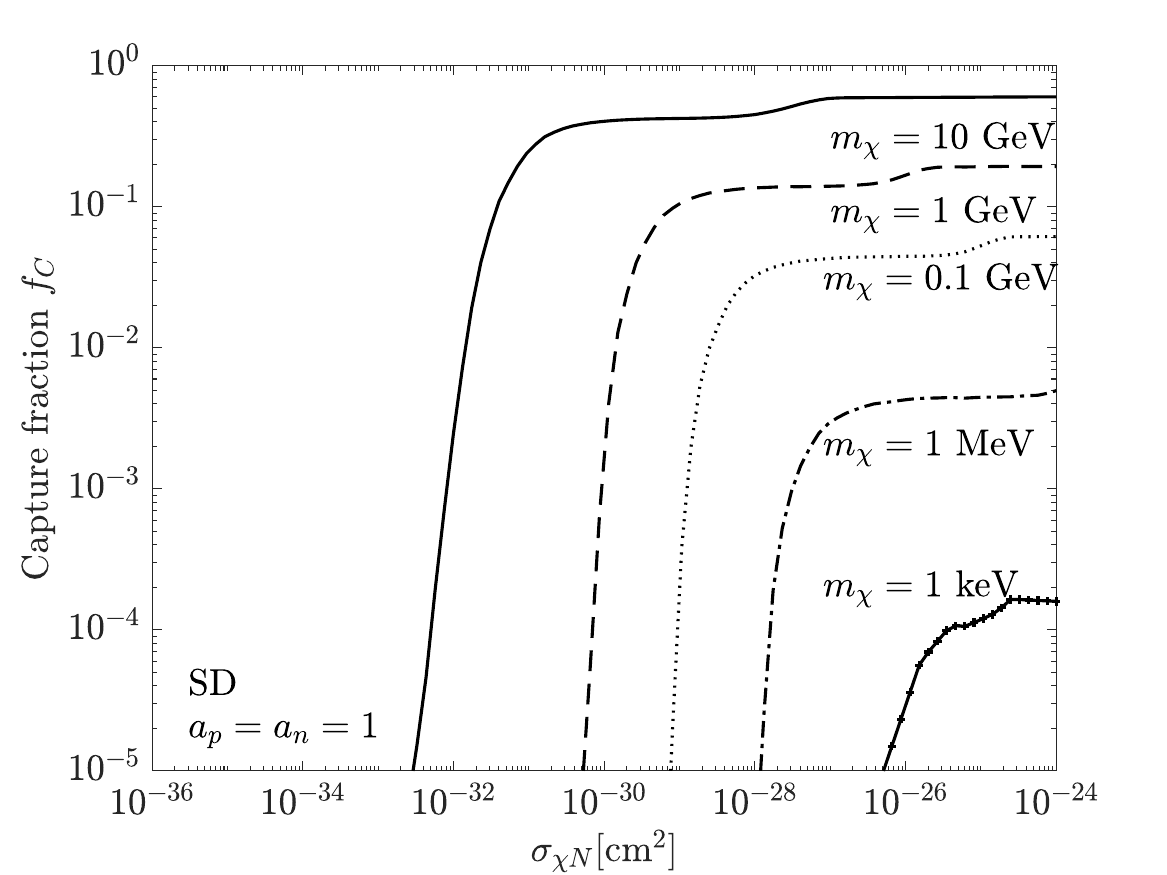}    
    \caption{Dark matter capture in the Earth for spin-independent interactions from Monte Carlo simulation using~\texttt{DaMaSCUS-EarthCapture}, assuming isospin-independent nuclear response $a_p=a_n=1$. {\it Left:} The fraction of dark matter particles that are captured among those impinge on the Earth as a function of dark matter mass. The orange, magenta, purple, green, light blue and yellow lines correspond to the cross section of $10^{-34}$, $10^{-32}$, $10^{-30}$,  $10^{-28}$, $10^{-26}$, $10^{-24}$~cm$^2$, respectively. {\it Right:} The capture fraction of dark matter as a function of dark matter-nucleon scattering cross section. From bottom to top, different line types delineate dark matter masses of $m_\chi=10$~GeV, 1~GeV, 100~MeV, 1~MeV, and 1~keV respectively.}
    \label{fig:capMC_SD}
\end{figure}

\begin{figure}
    \centering
    \includegraphics[width=0.48\textwidth]{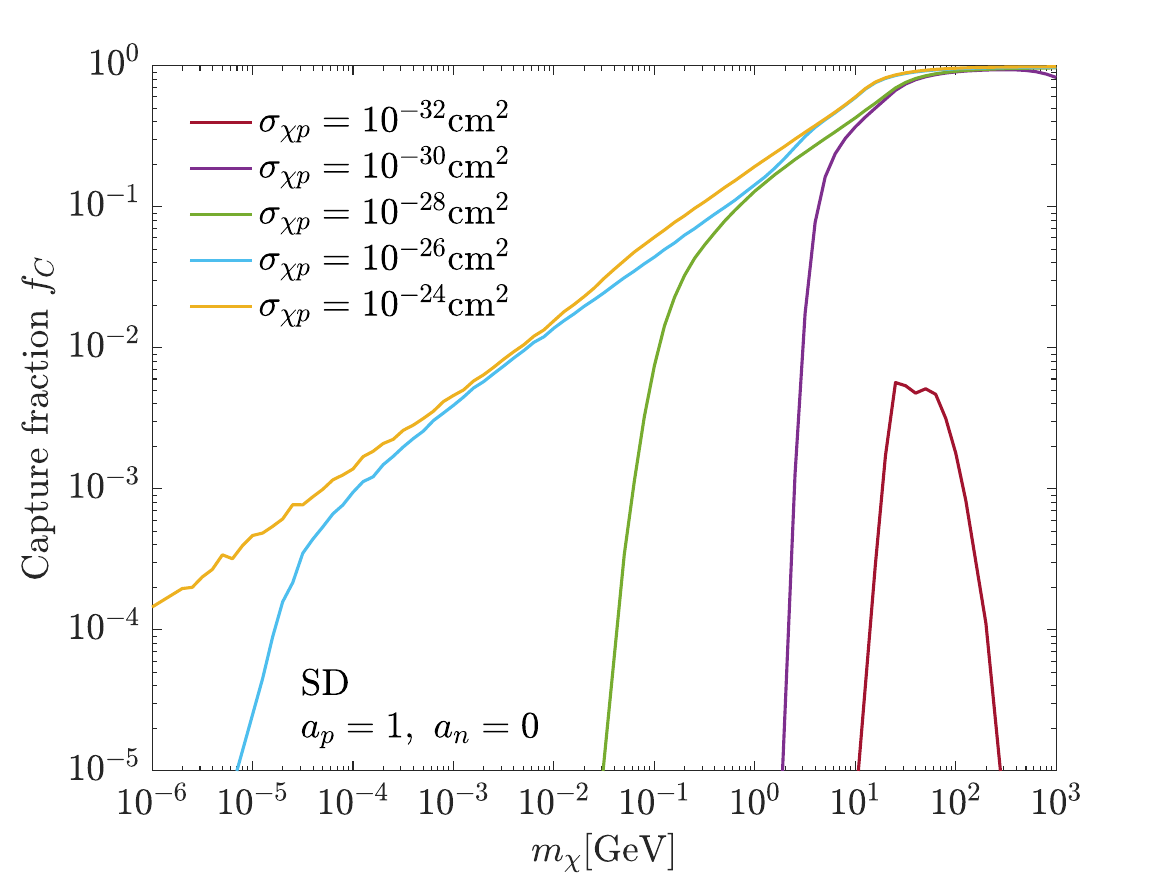}
    \includegraphics[width=0.48\textwidth]{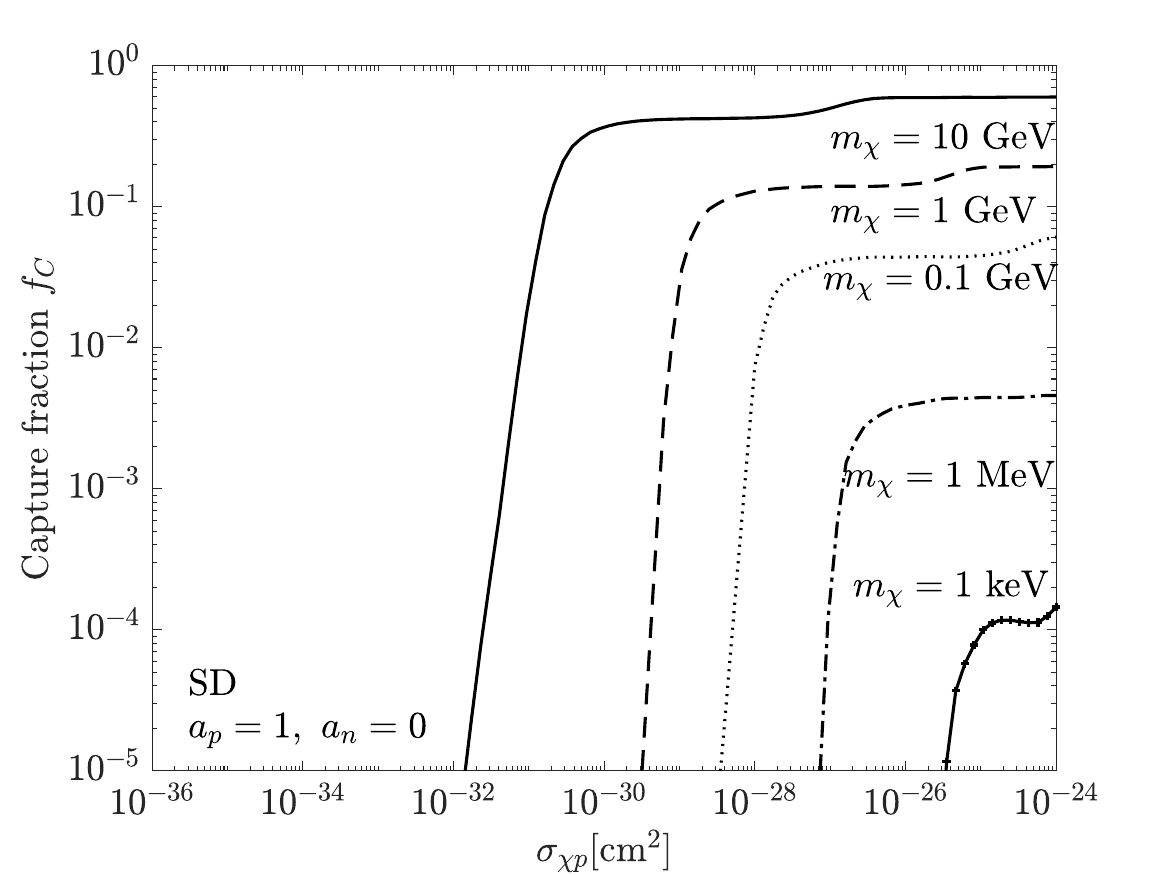}    
    \caption{Same as Figure~\ref{fig:capMC_SD} but for spin-dependent proton-only interaction.}
    \label{fig:capMC_SDp}
\end{figure}

\begin{figure}
    \centering
    \includegraphics[width=0.48\textwidth]{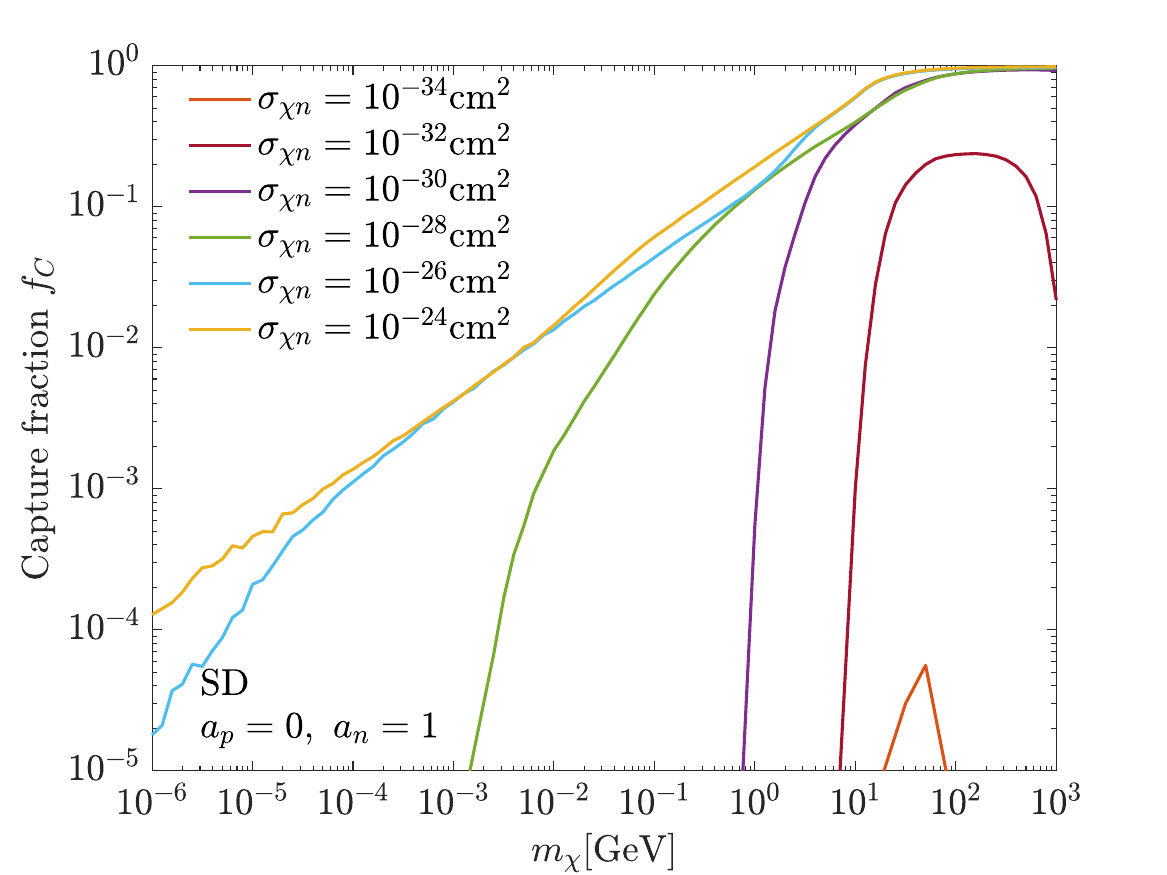}
    \includegraphics[width=0.48\textwidth]{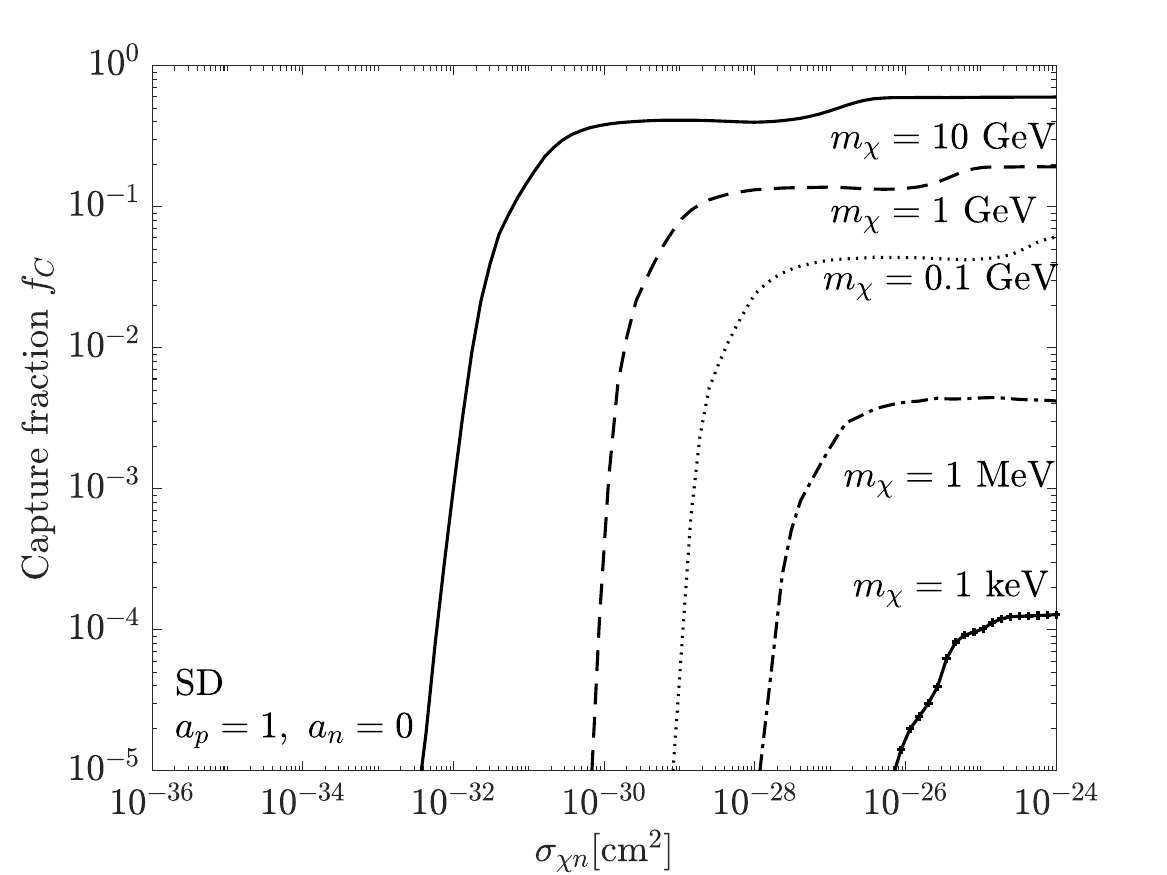}    
    \caption{Same as Figure~\ref{fig:capMC_SD} but for spin-dependent neutron-only interaction.}
    \label{fig:capMC_SDn}
\end{figure}

\begin{figure}
    \centering
    \includegraphics[width=0.6\textwidth]{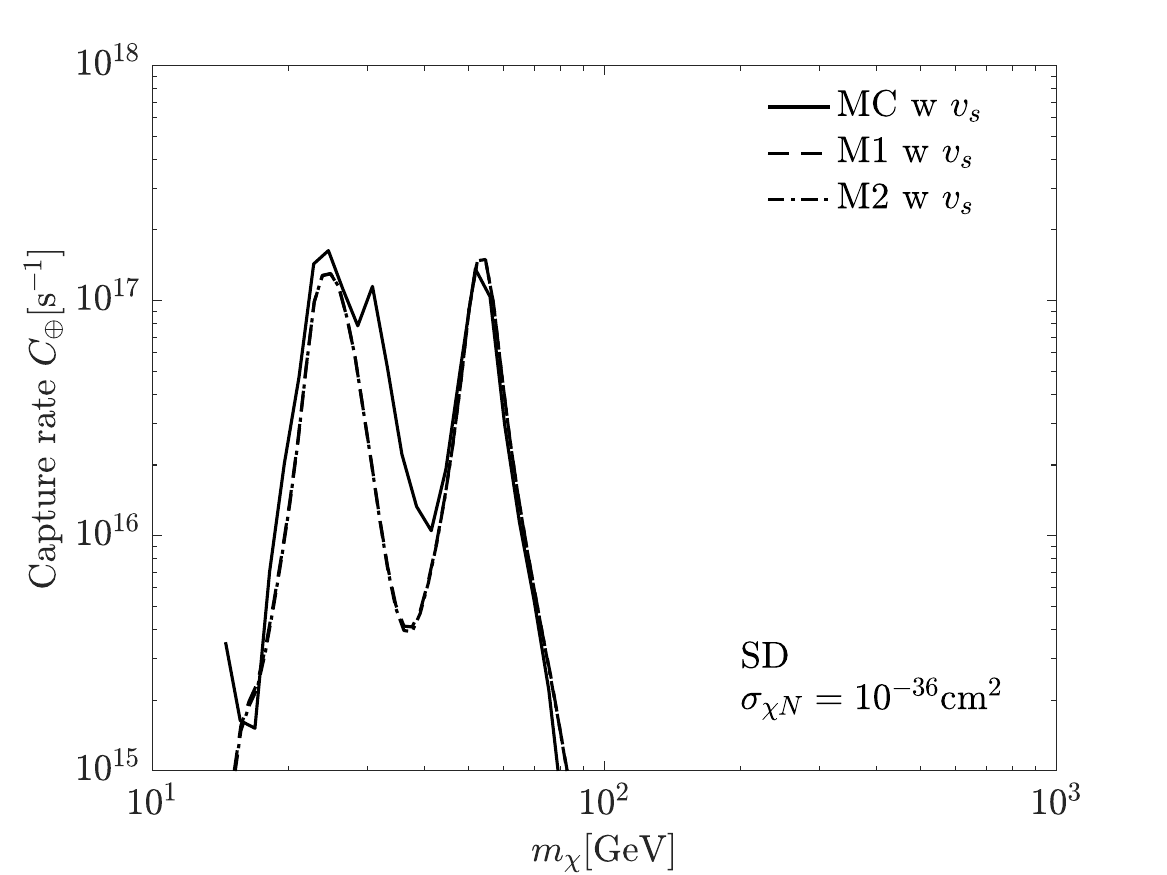}
    \caption{Dark matter capture rate in the Earth for spin-dependent scattering cross section $\sigma_{\chi N}^{\rm SD}=10^{-36}$~cm$^2$, $a_p=a_n=1$. The dashed, dash-dotted and solid lines show the capture rates from M1~\cite{Garani:2017jcj}, M2~\cite{Baum:2016oow} and from Monte Carlo simulations using~\texttt{DaMaSCUS-EarthCapture}, respectively.}
    \label{fig:comparecap}
\end{figure}

To give a complete picture of dark matter capture, we show the dark matter capture fraction for isospin-independent, proton-only and neutron-only spin dependent scatterings in Figure~\ref{fig:capMC_SD} to Figure~\ref{fig:capMC_SDn}. It is clearly seen from the left panel of Figure~\ref{fig:capMC_SD} that in the limit where single scatter contributes significantly to the capture ($\sigma_{\chi N}^{\rm SD}=10^{-34}$~cm$^2$), the capture fraction peaks between 10~GeV and 100~GeV, where the dark matter mass kinematically matches the masses of \Nuc{Si}{29}, \Nuc{Al}{27}, \Nuc{Mg}{25}, as well as \Nuc{Fe}{57}, and the dark matter kinetic energy loss is also maximized. A comparison between MC simulation and the single-scatter analytical results in the very low cross section limit is also shown in Figure~\ref{fig:comparecap}. The simulation matches the  analytical results well, apart from small difference attributable to limited statistics. However, at higher cross section, multiple scattering becomes important and the peaks are smeared out.

\section{Dark Matter Evaporation \label{app:DM_evap}}
The dark matter radial distribution assuming local thermal equilibrium follows~\cite{Garani:2017jcj}
\begin{equation}
    \dfrac{n_\chi(r)}{n_{\chi}(0)}=\left(\dfrac{T_\oplus(r)}{T_\oplus(0)}\right)^{3/2}\exp\left(-\int_0^r\left[\alpha (r')\frac{dT_\oplus(r')}{dr'}
    +m_\chi\frac{d\phi(r')}{dr'}\right]T_\oplus^{-1}dr'\right)\,,
\end{equation}
where $n_\chi(0)$ and $T_\oplus(0)$ are the dark matter density and Earth temperature, respectively, at $r=0$, 
and $\phi (r)$ is the external gravitational potential ($\phi (r) 
=\int_0^{r'}GM_\oplus(r')/r'^2dr'$).  The thermal 
diffusivity $\alpha (r)$ 
is related to the mean free path of dark matter in the Earth (see~\cite{Garani:2017jcj} for its explicit form). 
For velocity-independent interaction, $\alpha$ is independent of cross section and mildly dependent on the nature of the interaction.
A self-consistent study was recently carried out in Ref.~\cite{Leane:2022hkk} including the effects of diffusion and gravity, which can result in a floating distribution of dark matter on the Earth's surface. We have checked that this ``bouyant'' dark matter distribution does not change the overall distribution for masses above 1 GeV significantly and our results remain robust.

\begin{figure}
    \centering
    \includegraphics[width=0.48\textwidth]{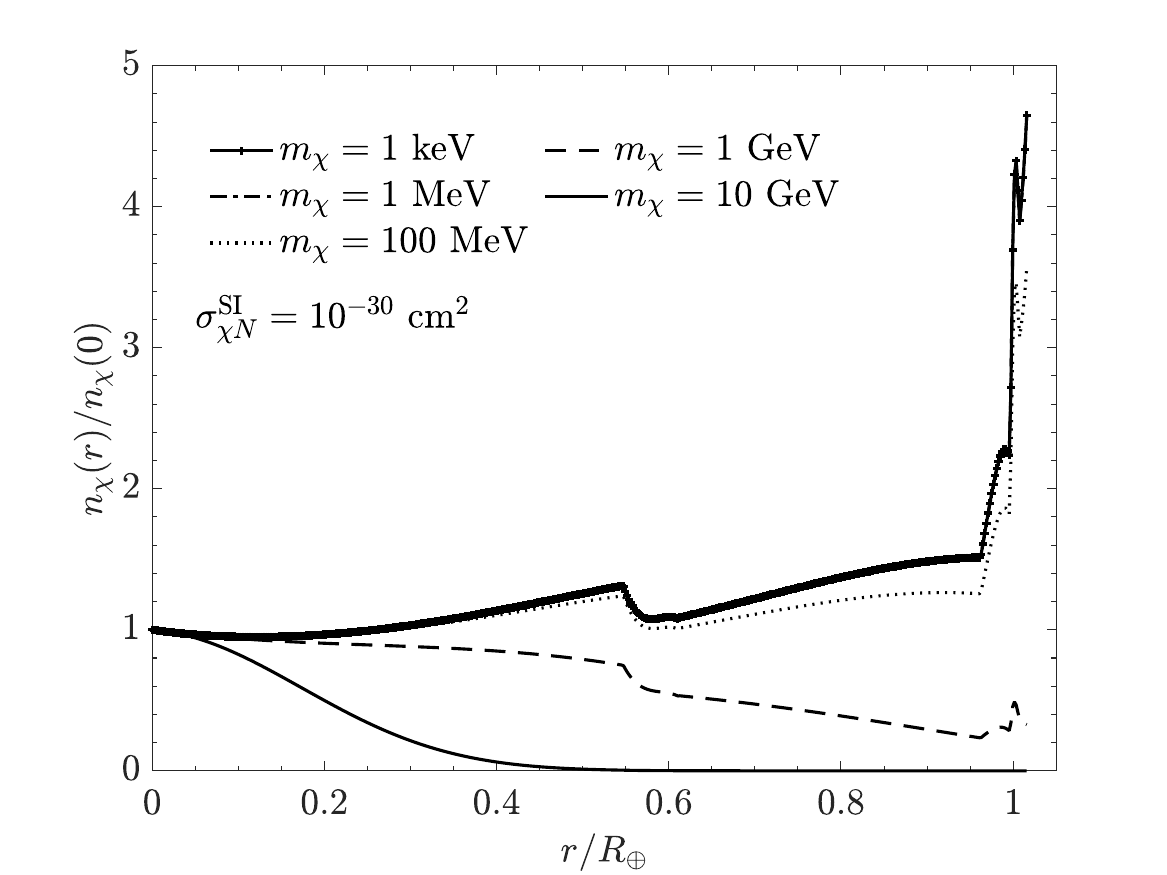}
    \includegraphics[width=0.48\textwidth]{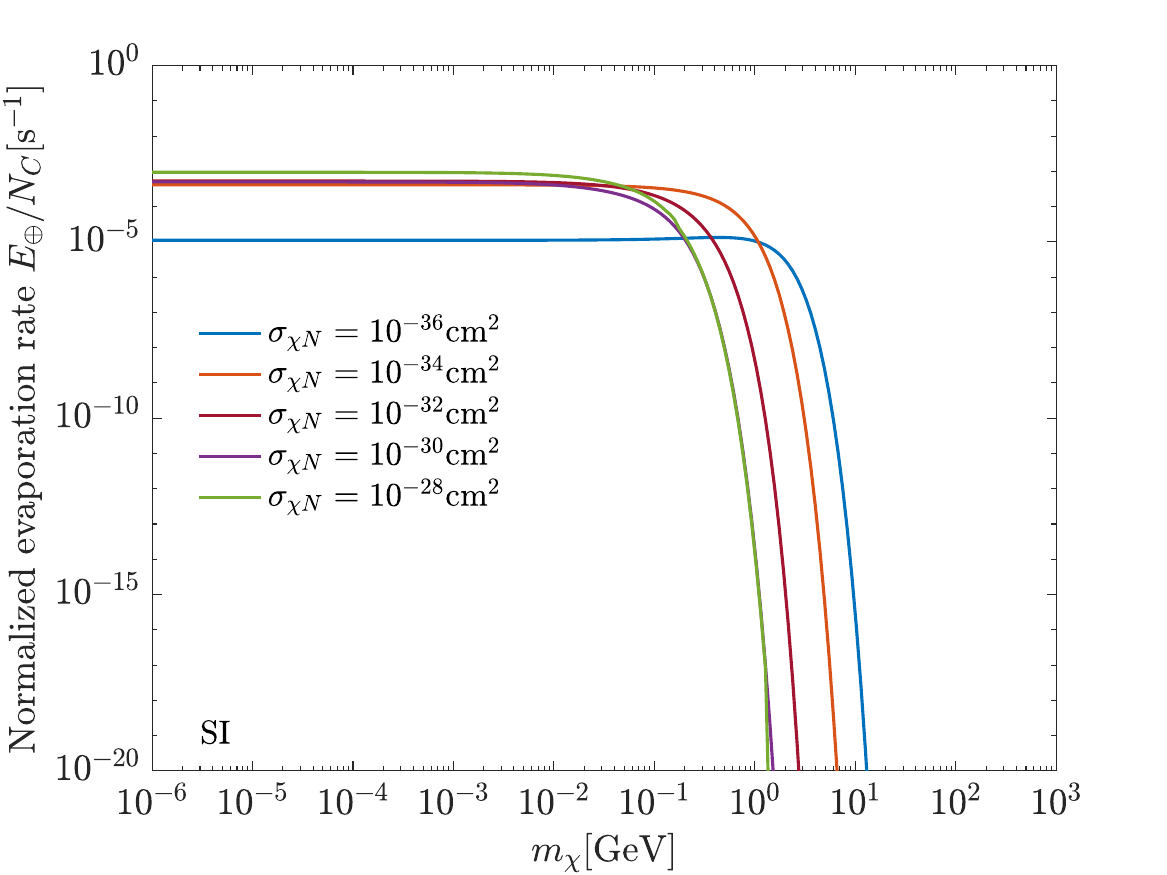}
    \caption{{\it Left:} Normalized dark matter number density distribution as a function of the distance from the center of the Earth including 100~km atmosphere, for a benchmark spin-independent cross-section of $10^{-30} \rm cm^{2}$. Line styles correspond to dark matter masses of 1~keV, 1~MeV, 100~MeV, 1~GeV, 100~GeV, separately, from top to bottom. {\it Right:} The normalized evaporation rate per captured particle $E_\oplus/N_C$ as a function of dark matter mass, for the various spin-independent dark matter-nucleon cross sections indicated.}
    \label{fig:ndist}
\end{figure}

The radial distribution of dark matter is depicted in the left panel of Figure~\ref{fig:ndist}. For relatively light dark matter $m_\chi\lesssim 0.1$~GeV, captured dark matter dwells towards the surface of the Earth due to the temperature gradient in the upper mantle and the crust. Due to the temperature profile in the atmosphere, the density distribution fluctuates at a radius $\RE<r\leq \Ratm$. Heavy dark matter particles tend to sink down. In particular, heavier dark matter $m_\chi\gtrsim 10$~GeV is more clustered in the Earth core.

Assuming local thermal equilibrium, the velocity of captured dark matter follows a Maxwell-Boltzmann distribution (at zeroth order in the temperature gradient), 
with a cutoff at the local escape velocity
\begin{equation}
    f_\oplus(\vec{u}_\chi,r)=\dfrac{1}{(\sqrt{\pi}v_\chi(r))^3}e^{-u_\chi^2/v_\chi^2(r)}\left({\rm Erf}(x)-\frac{2}{\sqrt{\pi}}xe^{-x^2}\right)^{-1}\Theta(v_e(r)-u_\chi)\,,
\end{equation}
where again $v_e(r)$ is the escape velocity at radius $r$. The dark matter thermal velocity is $v_\chi=\sqrt{2T_\oplus(r)/m_\chi}$ and we define $x\equiv v_e/v_\chi$. Due to the thermal motion of Earth nuclei, dark matter may scatter with a nucleus and acquires a high enough velocity to escape from the Earth.
The corresponding evaporation rate is described as~\cite{Garani:2017jcj}
\begin{equation}
    E_\oplus = \sum\limits_j\int_0^\RE 4\pi r^2 n_\chi(r)s(r)dr\times
    \int_0^{v_e(r)}4\pi u_\chi^2f_\oplus du_\chi\int_{v_e(r)}^\infty R_j^+(u_\chi\rightarrow v)dv\,.
    \label{eq:evaprates}
\end{equation}
Note that in the optically thick regime, dark matter particles with a velocity above the escape velocity may not actually evaporate, as they may scatter with the Earth matter several times before making their way out. This effect is encapsulated in the $s(r)$ factor defined as~\cite{Garani:2017jcj} $s(r)=\eta_{\rm ang}\eta_{\rm mult} e^{-\tau(r)}$, where
$\eta_{\rm ang}$ and $\eta_{\rm mult}$ take into account the angular trajectory and multiple scattering respectively, and $\tau(r)$ is the optical depth. 
Note that the only dark matter properties which $E_\oplus$ depends on are the dark matter mass, scattering cross section and 
the constant $n_\chi(0)$.

The dark matter evaporation rates as obtained from Eq.~\eqref{eq:evaprates} for spin-independent interaction are displayed in the right panel of Figure~\ref{fig:ndist}. As the mass increases, the evaporation rate is exponentially suppressed by the thermal velocity of dark matter. Dark matter with mass $m_\chi\gtrsim 10$~GeV can hardly escape from the Earth regardless of the cross section. Evaporation is enhanced when
$\sigma_{\chi N}^{\rm SI}$ goes above $10^{-36}$~cm$^2$, thanks to the rise of the scattering rate between dark matter and nuclei. However, further increasing the cross section to $10^{-34}$~cm$^2$ or higher will not facilitate more evaporation, as the $s(r)$ factor becomes important and dark matter from the inner layer could hardly find a way out of the Earth. In this scenario, the evaporation is more and more sourced from the dark matter particles near the Earth crust or even the atmosphere. The combination of the effects of more scattering and narrower evaporation region makes the evaporation of light dark matter rather insensitive to the cross section when $\sigma_{\chi N}\gtrsim 10^{-34}$~cm$^2$. Similar to the capture analysis, Eq.~\eqref{eq:evaprates} might overestimate the evaporation rate for $m_\chi\ll m_A$  while 
the direction of the dark matter particle is effectively randomized at every scatter when dark matter makes its way out. A dedicated Monte Carlo study is required to obtain the proper evaporation rate at the low mass regime, which we leave for future work. Our analysis using Eq.~\eqref{eq:evaprates} is conservative.

\section{Constraints from Direct Detection Experiments and Cosmology}
\label{app:Directdetection}

We also show constraints from CMB~\cite{Dvorkin:2013cea,Gluscevic:2017ywp}, XQC~\cite{Erickcek:2007jv}, RRS~\cite{rich1987search}, CRESST 2017 surface run~\cite{CRESST:2017ues}, CDMS-I~\cite{CDMS:2002moo}, CRESST-III~\cite{CRESST:2019jnq} and XENON1T~\cite{XENON:2018voc}. For the CRESST surface run we use the upper limit from Refs.~\cite{Kavanagh:2017cru,Davis:2017noy} and the lower limit from Ref.~\cite{CRESST:2017ues}. For CDMS-I we adopt the upper limit from~\cite{Kavanagh:2017cru} and the lower limit from~\cite{CDMS:2002moo}.   As RRS placed a constraint on the dark matter-silicon scattering cross section, we translate that to a dark matter-nucleon scattering cross section using the relations described in the main text. To determine bounds in this parameter space for for CRESST-III and XENON1T, we 
derive our limits by assuming that
dark matter particles all arrive from the average zenith angle of $54^\circ$, and do not change their direction of motion significantly as they pass through the Earth. The speed change of dark matter per unit distance is given by~\cite{Bhoonah:2020dzs,Kavanagh:2017cru}
\begin{equation}
    \dfrac{du_\chi}{dD}=-\dfrac{u_\chi}{m_\chi}\sum\limits_j \dfrac{\mu_{A_j}^2 n_j}{m_{A_j}}\sigma_j\,,
    \label{eq:dudD}
\end{equation}
where $m_{A_j}$, $\mu_{A_j}$ and $n_j$ are the nucleus mass, dark matter-nucleus reduced mass, and number density in the 
Earth, respectively,  of the $j$th isotope.  The final one-dimensional dark matter velocity at the detector is connected to the halo dark matter velocity distribution\footnote{For this purpose, 
we can ignore the effects of gravitational infall.  Particles which are slow enough that gravitational effects are important will in any case 
not deposit enough energy in the detector to exceed threshold.} 
via
\begin{equation}
    f(u_f)=f(u_\chi)\dfrac{du_\chi}{du_f}=f(u_\chi)\exp\left(\int dD \sum\limits_j \dfrac{\mu_{A_j}^2 n_j}{m_\chi m_{A_j}}\sigma_j\right)\,.
    \label{eq:fuf}
\end{equation}
For simplicity we do not include form factors in the overburden calculation using Eq.~\eqref{eq:fuf}, \ie~$\sigma_j\simeq \sigma_{j,0}$. The expected number of events in an experiment is\footnote{Note that we do not include the momentum transfer-dependent form factors in overburden calculations for simplicity, but we always include form factors in the terrestrial experiments to produce as accurate experimental limits as possible.}
\begin{equation}
    N_{\rm exp}=\sum\limits_i N_i T \dfrac{\rho_\chi}{m_\chi}\int u_ff(u_f)du_f\int \dfrac{d\sigma_i}{dE_R}\epsilon(E_R)dE_R\,,
    \label{eq:Nexp}
\end{equation}
where
\begin{equation}
    \dfrac{d\sigma_i}{dE_R}=\dfrac{\sigma_{i,0}m_{A_i}}{2\mu_{A_i}^2u_f^2}F_i^2(E_R)\,,
    \label{eq:dsigmadE}
\end{equation}
$N_i$ is the number of target nuclei, and $T$ is the exposure time. We also include the efficiency factor $\epsilon(E_R)$ from the respective experiments.
For spin-independent interactions, $F_i$ is the Helm form factor. We use Eq.~\eqref{eq:Nexp} to find the cross section limits for CRESST-III and XENON1T. 441 dark matter candidate events were identified with the exposure of 3.64~kg$\cdot$days 
using CaWO$_4$ crystal~\cite{CRESST:2019jnq}, 
while for XENON1T we adopt 3.7 events at $90\%$ upper limit with 0.9 tonne reference mass~\cite{XENON:2018voc}.  

 We also derive the constraints when $\chi$ constitutes $5\%$ dark matter. 
To produce these limits, we scale up the lower limit of XQC and RRS by a factor of 20. The same scaling relation also applied to the CMB limit, as the collision terms enters the Boltzmann equations in the form of $\rho_\chi\sigma_{\chi p}$~\cite{Gluscevic:2017ywp}. We assume the overburden line remains robust with reduced dark matter flux for XQC and RRS. We use the \texttt{verne}~\cite{verne,Kavanagh:2017cru} code to compute the dark matter limits from CRESST surface run and CDMS-I, where the dark matter incoming angle and velocity distribution are taken into consideration in the overburden calculation.  We again use Eq.~\eqref{eq:Nexp} to find the constraints from CRESST-III and XENON1T, when $N_{\exp}$ matches the observation limit 
with a reduced dark matter 
density.

For spin-dependent nuclear scattering, we follow \crefrange{eq:dudD}{eq:dsigmadE} to derive the constraints from XENON1T and CDMSlite~\cite{SuperCDMS:2017nns}. The spin-dependent form factor $F^2_{\rm SD}=(a_p S_p(E_R)+a_n S_n(E_R))/S_T(0)$, where
\begin{equation}
    S_T(0)=\dfrac{(2J+1)(J+1)}{\pi J}\left(a_p\langle S_p\rangle+a_n\langle S_n\rangle \right)^2\,.
\end{equation}
We employ the updated nuclear structure factors and average spins in~\cite{Klos:2013rwa}, which are also listed in Table~\ref{tab:nuclearspins}.
We use the first two energy bins in CDMSlite Run 2, which well reproduces the results in~\cite{SuperCDMS:2017nns}. We again use \texttt{verne} to compute the limits from CRESST surface run and CDMS-I, neglecting the form factors in the overburden, but including them in the scattering rates in the detector, except for \Nuc{O}{17} whose momentum dependent structure factor is still missing. The XQC and RRS limits are translated to the nucleon scattering cross section accrodingly. For the overburden of these two experiments we assume \Nuc{N}{14} is the main source of dark matter deceleration in both spin-dependent and spin-independent scattering, and translate the constraints accordingly. For the case in which $\chi$ is 5\% of cosmological dark matter we follow the same treatment as in the spin-independent scattering case.

\clearpage

\bibliography{evap.bib}

\end{document}